\documentclass{ar-1col}
\usepackage[numbers]{natbib}
\usepackage{url}
\usepackage{bm}
\usepackage{amsmath}
\DeclareMathOperator{\Tr}{Tr}
\jname{Xxxx. Xxx. Xxx. Xxx.}
\jvol{AA}
\jyear{YYYY}
\doi{10.1146/((please add article doi))}

\begin{document}

\markboth{Selinger}{Director Deformations, Geometric Frustration, and Modulated Phases}

\title{Director Deformations, Geometric Frustration, and Modulated Phases in Liquid Crystals}

\author{Jonathan V. Selinger
\affil{Department of Physics, Advanced Materials and Liquid Crystal Institute, Kent State University, Kent, Ohio 44242, USA; email: jselinge@kent.edu}}

\begin{abstract}
This article analyzes modulated phases in liquid crystals, from the long-established cholesteric and blue phases to the recently discovered twist-bend, splay-bend, and splay nematic phases, as well as the twist-grain-boundary (TGB) and helical nanofilament variations on smectic phases.  The analysis uses the concept of four fundamental modes of director deformation:  twist, bend, splay, and a fourth mode related to saddle-splay.  Each mode is coupled to a specific type of molecular order:  chirality, polarization perpendicular and parallel to the director, and octupolar order.  When the liquid crystal develops one type of spontaneous order, the ideal local structure becomes nonuniform, with the corresponding director deformation.  In general, the ideal local structure is frustrated; it cannot fill space.  As a result, the liquid crystal must form a complex global phase, which may have a combination of deformation modes, and may have a periodic array of defects.  Thus, the concept of an ideal local structure under geometric frustration provides a unified framework to understand the wide variety of modulated phases.
\end{abstract}

\maketitle

\tableofcontents

\section{INTRODUCTION}

Liquid crystals exhibit a wide variety of modulated phases, in which the director field $\hat{\bm{n}}(\bm{r})$ has some periodic variation as a function of position.  Some of these modulated phases, the cholesteric and blue phases, have been known since the earliest days of liquid crystal research~\cite{Dunmur2011}.  Other modulated phases have only been discovered recently, as discussed below.  In this review article, we suggest a unified way of thinking about all of these modulated phases.  Our argument is based on two general principles:

First, we consider the four fundamental deformation modes of a nematic director field---the well-known bend, splay, and twist deformations, and a less-well-known fourth mode related to saddle-splay---which have been identified by a recent theoretical analysis~\cite{Machon2016,Selinger2018}.  Each of these modes couples with some type of molecular order, in addition to the standard nematic orientational order.  This coupling can be regarded as ``generalized flexoelectricity.''  Each type of molecular order might form spontaneously, to make a phase with nematic order and a small amount of extra order, which we will call an ``$N+\epsilon$'' phase.  In such a phase, the extra order induces the corresponding deformation of the nematic director field.  The most common case is that chiral order induces spontaneous twist, but the other three cases are also possible, and they lead to other types of ideal nonuniform local structures.

Second, we consider the concept of geometric frustration, which describes a system with an ideal local structure that cannot fill space because it violates some global constraint.  One simple example is an Ising antiferromagnet on a triangular lattice, where the ideal local alternation between spin-up and spin-down is incompatible with the three-fold symmetry of the lattice.  This concept is widely used in the study of magnetic systems, where it explains the formation of complex modulated phases~\cite{Moessner2006}.  It is also used in the study of soft materials, where it provides a mechanism to limit the size of self-assembled aggregates~\cite{Grason2016}.   In the context of liquid crystals, geometric frustration means that the ideal local director deformation cannot fill space.  As a result, the liquid crystal forms a complex global phase that can fill space.  This global phase might have a mixture of the fundamental deformation modes, and it might have a periodic array of defects.

\begin{table}
\caption{Director deformations, corresponding molecular order, and achievable phases}
\label{tab:modes}
\begin{center}
\begin{tabular}{@{}l|c|c|c|c|c@{}}
\hline
Deformation & Deformation & Mathematical & Corresponding & Achievable global & Achievable global \\
& considering $\hat{\bm{n}}\leftrightarrow-\hat{\bm{n}}$ & object & molecular order & phases (nematic) & phases (smectic-A) \\
\hline
Twist & Twist & Pseudoscalar & Chiral order & Cholesteric phase & Twist grain boundary\\
$T=\hat{\bm{n}}\cdot(\bm{\nabla}\times\hat{\bm{n}})$ & $T=\hat{\bm{n}}\cdot(\bm{\nabla}\times\hat{\bm{n}})$ & & parameter & Blue phases & Helical nanofilaments\\
\hline
Bend & Bend & Vector & Polarization & Twist-bend $N_{TB}$ & Possible lattice of \\
$\bm{B}=\hat{\bm{n}}\times(\bm{\nabla}\times\hat{\bm{n}})$ & $\bm{B}=\hat{\bm{n}}\times(\bm{\nabla}\times\hat{\bm{n}})$ & $\perp$ to $\hat{\bm{n}}$ & $\bm{P}_\perp$ & Splay-bend $N_{SB}$ & edge dislocations \\
\hline
Splay scalar & Splay vector & Vector & Polarization & Splay nematic $N_S$ & Spherical shells \\
$S=\bm{\nabla}\cdot\hat{\bm{n}}$ & $\bm{S}=S\hat{\bm{n}}=\hat{\bm{n}}(\bm{\nabla}\cdot\hat{\bm{n}})$ & $\parallel$ to $\hat{\bm{n}}$ & $\bm{P}_\parallel$ & \\
\hline
$\Delta_{ij}$ & $\Delta_{ij}n_k$ & Tensor of & Octupolar order & Not yet studied & Helical nanofilaments\\
& & rank 3* & parameter* & \\
\hline
\end{tabular}
\end{center}
\begin{tabnote}
*First two legs traceless, symmetric, perpendicular to $\hat{\bm{n}}$, third leg parallel to $\hat{\bm{n}}$.
\end{tabnote}
\end{table}

In the following sections, we show how this analysis applies to four types of molecular order, which couple to the four fundamental director deformation modes and induce modulated phases.  We first consider nematic liquid crystals, in which the global constraints arise merely from the geometry of three-dimensional (3D) Euclidean space.  We then go on to smectic liquid crystals and other layered structures, in which the layering gives more severe global constraints.  Most of those results are summarized in Table~\ref{tab:modes}.  Finally, we briefly discuss liquid crystal elastomers, where even more constraints come from the coupling between nematic order and the elasticity of a polymer network.

The current article can be regarded as a follow-up to a previous review article on ``Order and frustration in chiral liquid crystals'' \cite{Kamien2001}, which we coauthored with Kamien in 2001.  The single greatest difference is that the previous article concentrated on chirality as the mechanism to induce modulated phases, and the current article regards chirality as just one of four fundamental mechanisms.  Indeed, this difference shows progress in the field over twenty years, which now recognizes an even richer variety of complex modulated phases.

\section{NEMATIC LIQUID CRYSTALS}

\subsection{Director deformation modes}

In a nematic liquid crystal, the molecules have orientational order along a local axis, which is represented by the director field $\hat{\bm{n}}(\bm{r})$, but they do not have positional order.  To analyze director deformations, we use a mathematical approach that was recently developed for purposes of topology by Machon and Alexander~\cite{Machon2016}, and was applied to elasticity theory by Selinger~\cite{Selinger2018}.  This approach decomposes the director gradient tensor as
\begin{equation}
\partial_i n_j = -n_i B_j + \frac{1}{2}S(\delta_{ij}-n_i n_j) + \frac{1}{2}T\epsilon_{ijk}n_k + \Delta_{ij},
\end{equation}
with the four modes $\bm{B}$, $S$, $T$, and $\Delta_{ij}$ defined below.  In mathematical terminology, these modes are four irreducible representations of the group of rotations about the director $\hat{\bm{n}}$.

\begin{figure}
\tabcolsep0pt
\begin{tabular}{@{}cccc@{}}
(a) Bend $\bm{B}$ & (b) Splay $S$ & (c) Twist $T$ & (d) $\bm{\Delta}$ mode \\
\resizebox{!}{2.0cm}{\includegraphics{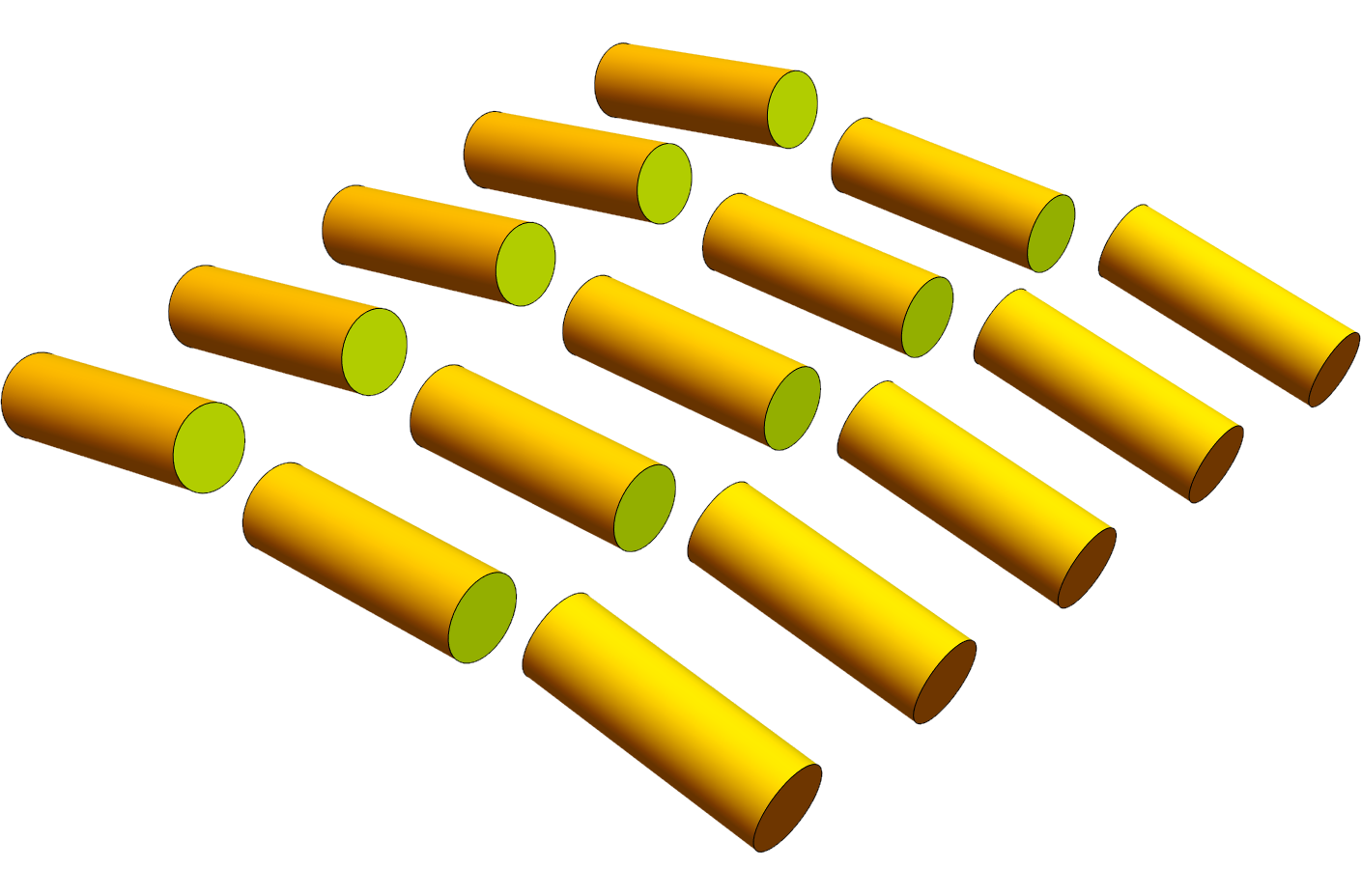}} &
\resizebox{!}{2.0cm}{\includegraphics{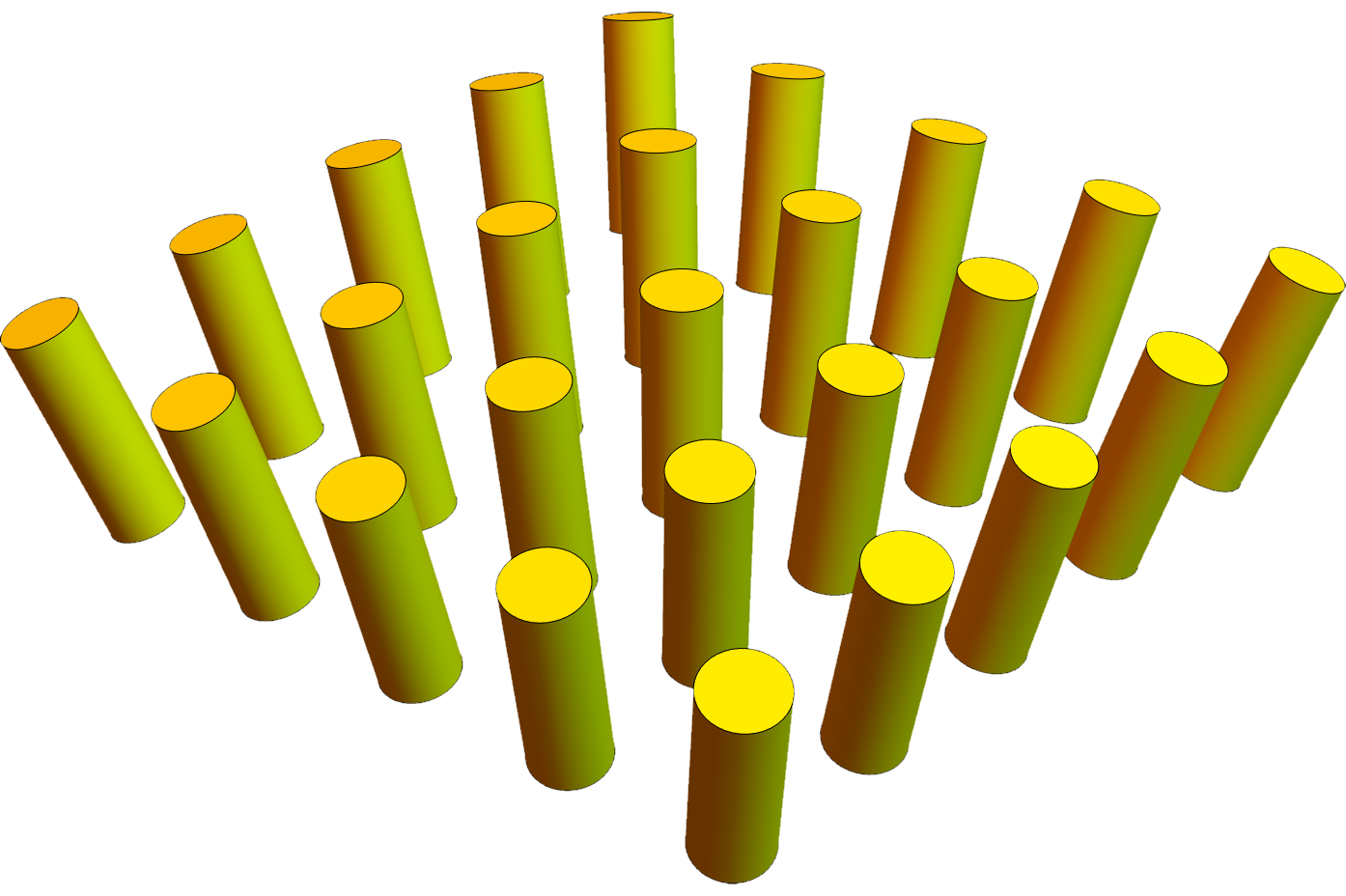}} &
\resizebox{!}{2.0cm}{\includegraphics{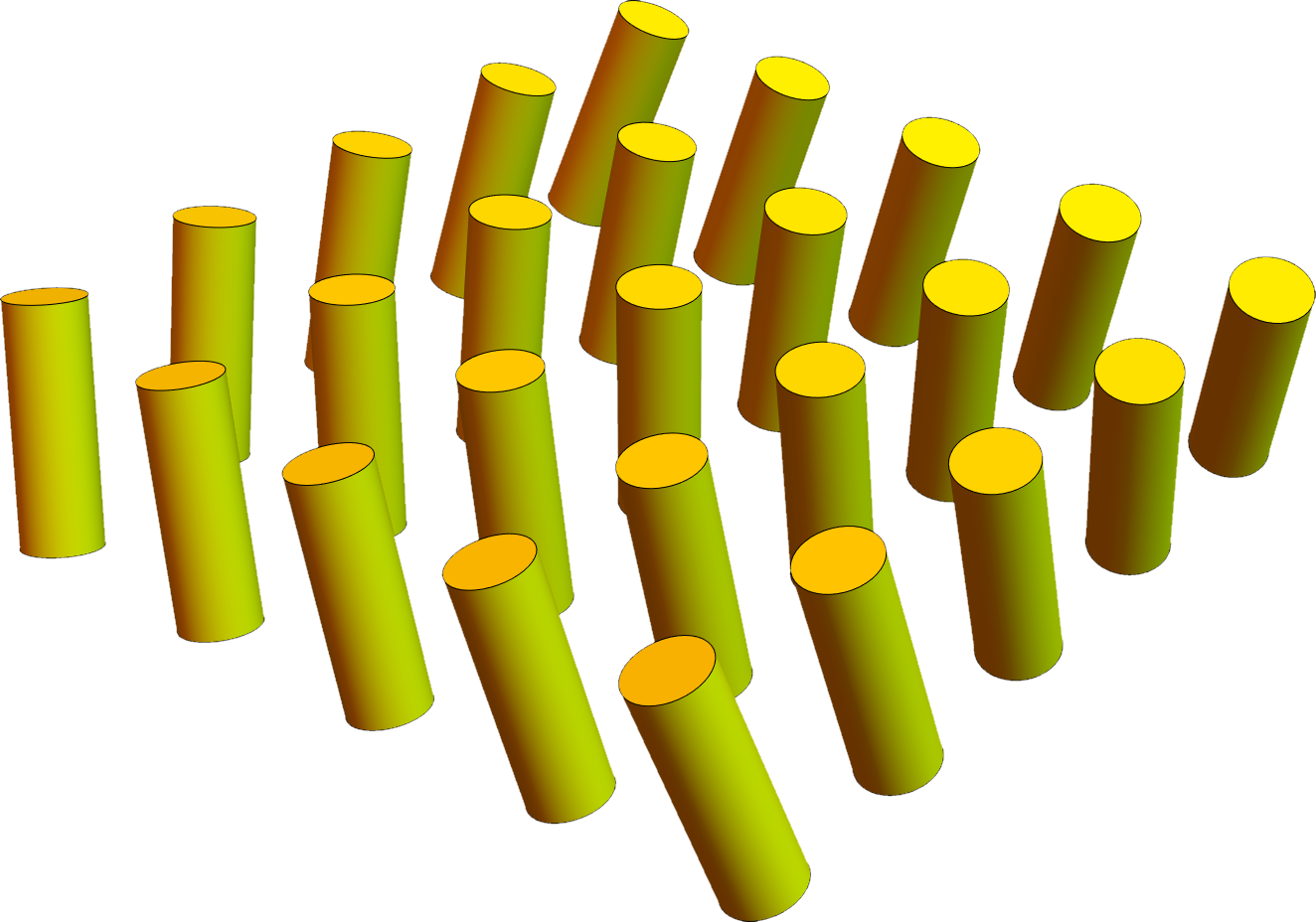}} &
\resizebox{!}{2.0cm}{\includegraphics{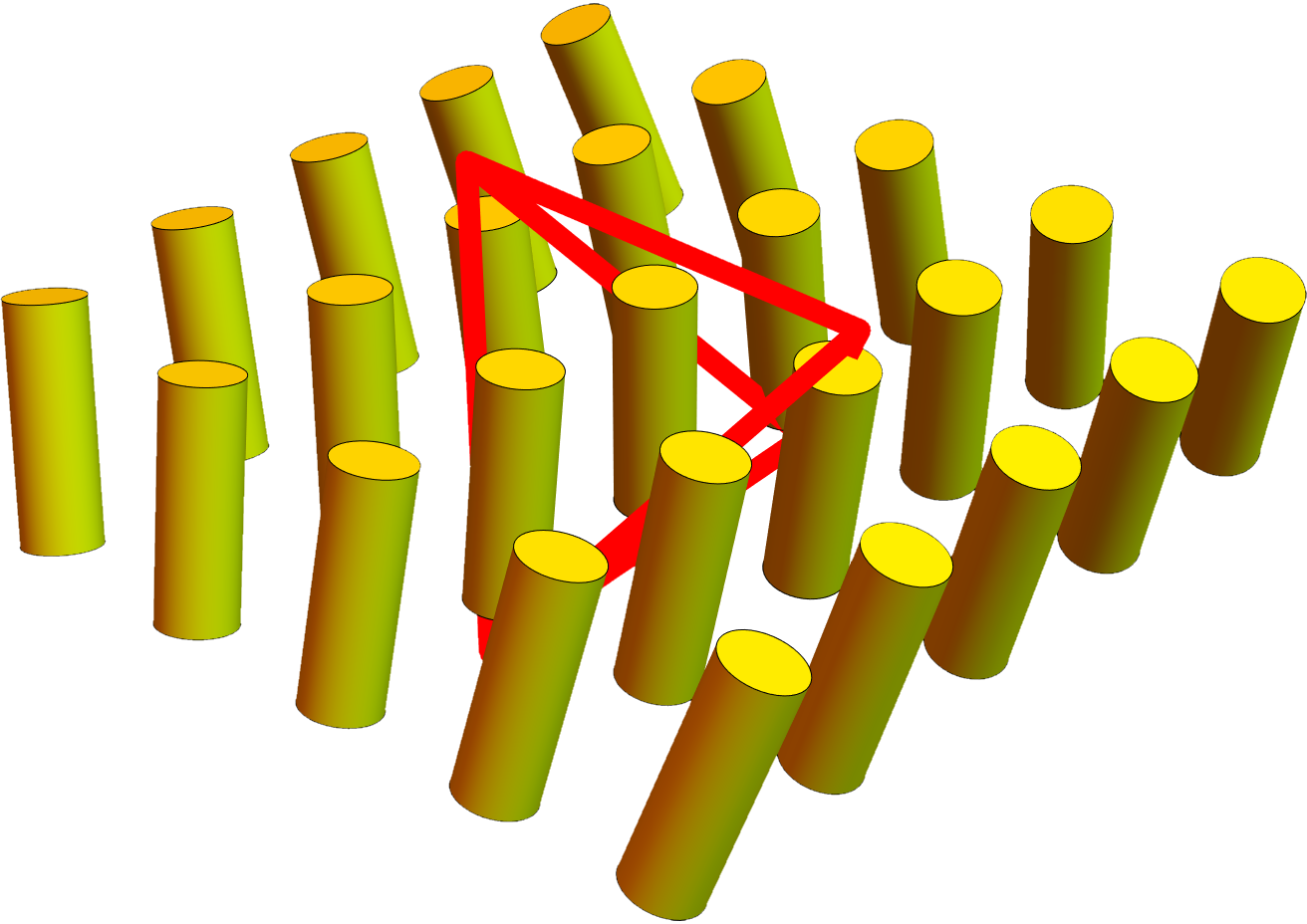}} \\
\end{tabular}
\caption{Director deformation modes in a nematic liquid crystal.}
\label{fig:modes}
\end{figure}

\subsubsection{Bend}  The first mode is bend, defined as $\bm{B}=\hat{\bm{n}}\times(\bm{\nabla}\times\hat{\bm{n}})=-(\hat{\bm{n}}\cdot\bm{\nabla})\hat{\bm{n}}$.  A deformation with locally pure bend, and all other modes equal to zero, is shown in Figure~\ref{fig:modes}(a).  This mode indicates how $\hat{\bm{n}}$ varies as one moves along the $\hat{\bm{n}}$ direction.  It is a vector in the plane perpendicular to $\hat{\bm{n}}$.  It has two degrees of freedom, because $\hat{\bm{n}}$ may change in either of the two directions perpendicular to $\hat{\bm{n}}$.  In a nematic liquid crystal, the director $\hat{\bm{n}}$ and $-\hat{\bm{n}}$ represent the same physical state.  Because the bend is even in $\hat{\bm{n}}$, it is invariant under the transformation $\hat{\bm{n}}\leftrightarrow-\hat{\bm{n}}$, and hence it is a uniquely defined physical object.

\subsubsection{Splay}  The second mode is splay, defined as $S=\bm{\nabla}\cdot\hat{\bm{n}}$.  A deformation with locally pure splay, and all other modes equal to zero, is shown in Figure~\ref{fig:modes}(b).  This mode is a scalar, with one degree of freedom.  It indicates how $\hat{\bm{n}}$ tilts inward or outward, isotropically in the plane perpendicular to $\hat{\bm{n}}$.  We emphasize that pure splay is double splay, in both directions perpendicular to $\hat{\bm{n}}$.  It is not single splay, like slices of pizza.  Indeed, it must be double splay because it is a scalar, which has no characteristic direction in the plane perpendicular to $\hat{\bm{n}}$.
Because $S$ is odd in $\hat{\bm{n}}$, it changes sign under the transformation $\hat{\bm{n}}\leftrightarrow-\hat{\bm{n}}$.  If we want a uniquely defined physical object, we must construct the splay vector $\bm{S}=S\hat{\bm{n}}=\hat{\bm{n}}(\bm{\nabla}\cdot\hat{\bm{n}})$, which is a vector parallel to $\hat{\bm{n}}$, still with one degree of freedom.

\subsubsection{Twist}  The third mode is twist, defined as $T=\hat{\bm{n}}\cdot(\bm{\nabla}\times\hat{\bm{n}})$.  A deformation with locally pure twist, and all other modes equal to zero, is shown in Figure~\ref{fig:modes}(c).  This mode is a pseudoscalar, with one degree of freedom.  It indicates how $\hat{\bm{n}}$ changes in a right- or left-handed way, isotropically in the plane perpendicular to $\hat{\bm{n}}$.  We emphasize that pure twist is double twist, in both directions perpendicular to $\hat{\bm{n}}$.  It is not single twist along a particular axis, like a cholesteric phase, which will be discussed later.  Indeed, it must be double twist because it is a pseudoscalar, which has no characteristic direction in the plane perpendicular to $\hat{\bm{n}}$.
Because $T$ is even in $\hat{\bm{n}}$, it is invariant under the transformation $\hat{\bm{n}}\leftrightarrow-\hat{\bm{n}}$, and hence it is a uniquely defined physical object.

\subsubsection{$\bm{\Delta}$ mode}  The fourth mode is $\Delta_{ij} = \partial_i n_j + n_i B_j - \frac{1}{2}S(\delta_{ij}-n_i n_j) - \frac{1}{2}T\epsilon_{ijk}n_k$.
A deformation with locally pure $\Delta_{ij}$, and all other modes equal to zero, is shown in Figure~\ref{fig:modes}(d).  This mode indicates how $\hat{\bm{n}}$ tilts outward in one direction, and inward in the orthogonal direction, in the plane perpendicular to $\hat{\bm{n}}$.  It is represented by a symmetric, traceless, second-rank tensor, in the plane perpendicular to $\hat{\bm{n}}$, which has two degrees of freedom.  Because $\Delta_{ij}$ is odd in $\hat{\bm{n}}$, it changes sign under the transformation $\hat{\bm{n}}\leftrightarrow-\hat{\bm{n}}$.  If we want a uniquely defined physical object, we must construct the third-rank tensor $\Delta_{ij}n_k$, still with two degrees of freedom.  When we include the equivalence between $\pm\hat{\bm{n}}$, we can see that this deformation has the symmetry of a tetrahedron, which splays outward in one direction with respect to $+\hat{\bm{n}}$, and outward in the orthogonal direction with respect to $-\hat{\bm{n}}$.  In Figure~\ref{fig:modes}(d), this tetrahedron is illustrated in red.

The terminology for this mode is not well established.  It was first defined by Machon and Alexander~\cite{Machon2016}, who called it ``anisotropic orthogonal gradients of $\hat{\bm{n}}$.''  That phrase is certainly accurate, but it seems rather long.  Selinger~\cite{Selinger2018} suggested the term ``biaxial splay,'' because it involves opposite splay in the two directions orthogonal to $\hat{\bm{n}}$.  However, when we consider the equivalence between $\pm\hat{\bm{n}}$, we can see that an even better term might be ``tetrahedral splay,'' which represents the full symmetry of the deformation.  For the rest of this article, we will just use the label ``$\bm{\Delta}$ mode,'' rather than choosing one of those terms.

\subsubsection{Elastic free energy}  The Oseen-Frank free energy density for deformations of the nematic director is conventionally written as
\begin{equation}
F_\text{nem}=\frac{1}{2}K_{11}S^2 + \frac{1}{2}K_{22}T^2 + \frac{1}{2}K_{33}|\bm{B}|^2
-K_{24}\bm{\nabla}\cdot\left[\hat{\bm{n}}(\bm{\nabla}\cdot\hat{\bm{n}})+\hat{\bm{n}}\times(\bm{\nabla}\times\hat{\bm{n}})\right],
\label{Fnemconventional}
\end{equation}
where the last term is called ``saddle-splay.''  Using the four modes defined above, the saddle-splay term can be rewritten as $-K_{24}[\frac{1}{2}S^2 + \frac{1}{2}T^2 - \Tr(\bm{\Delta}^2)]$, where $\Tr(\bm{\Delta}^2)=\Delta_{ij}\Delta_{ji}$.  Hence, an alternative form for the Oseen-Frank free energy density is
\begin{equation}
F_\text{nem}=\frac{1}{2}(K_{11}-K_{24})S^2 + \frac{1}{2}(K_{22}-K_{24})T^2 + \frac{1}{2}K_{33}|\bm{B}|^2
+ K_{24}\Tr(\bm{\Delta}^2).
\label{Fnem}
\end{equation}
That form is particularly convenient because it is a sum of squares, which shows that all four deformation modes cost elastic free energy.  Bend has elastic constant $K_{33}$, (double) splay has elastic constant $(K_{11}-K_{24})$, (double) twist has elastic constant $(K_{22}-K_{24})$, and the $\bm{\Delta}$ mode has elastic constant $2K_{24}$.  Reference~\cite{Selinger2018} provides several examples where this form of the free energy provides a new interpretation of older results in liquid-crystal physics.  We will use this form for the rest of this article.  We will not use the standard argument that the saddle-splay term in Equation~\ref{Fnemconventional} integrates to a surface term, because we are interested in phases with defects, which can be regarded as internal surfaces.

\subsection{Generalized flexoelectricity}

Each of the four deformation modes can induce molecular order that has the same symmetry as the deformation.  For bend and splay, this induced order was first predicted from symmetry considerations by Meyer~\cite{Meyer1969}, and it is now commonly observed as the flexoelectric effect.  For twist and the $\bm{\Delta}$ mode, we will consider this effect as ``generalized flexoelectricity.''

\begin{figure}
\tabcolsep0pt
\begin{tabular}{@{}cccc@{}}
(a) Uniform nematic & (b) Bent & (c) Uniform nematic & (d) Splayed \\
\rule[-1em]{0pt}{1em}%
\resizebox{!}{2.6cm}{\includegraphics{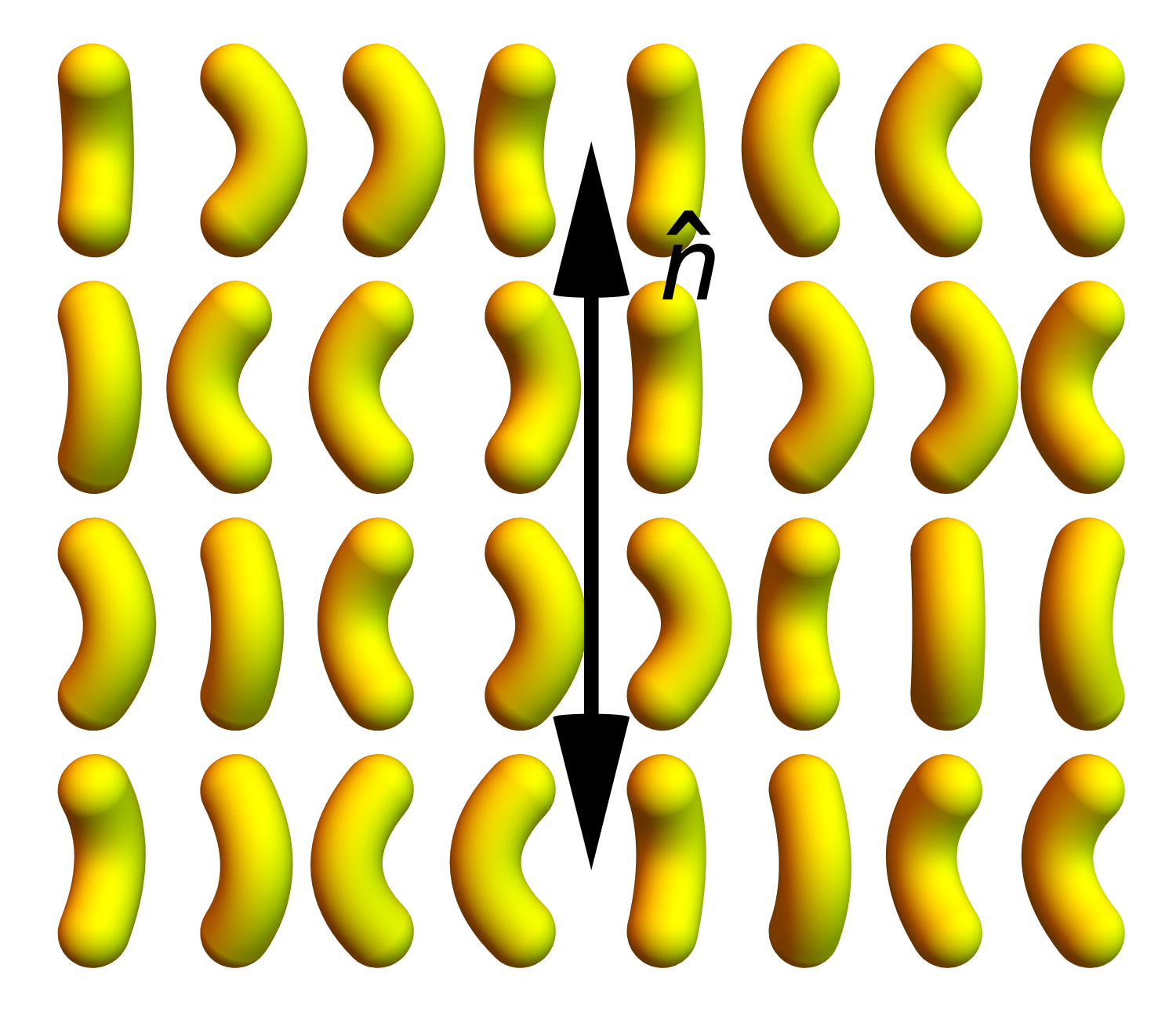}} &
\resizebox{!}{2.6cm}{\includegraphics{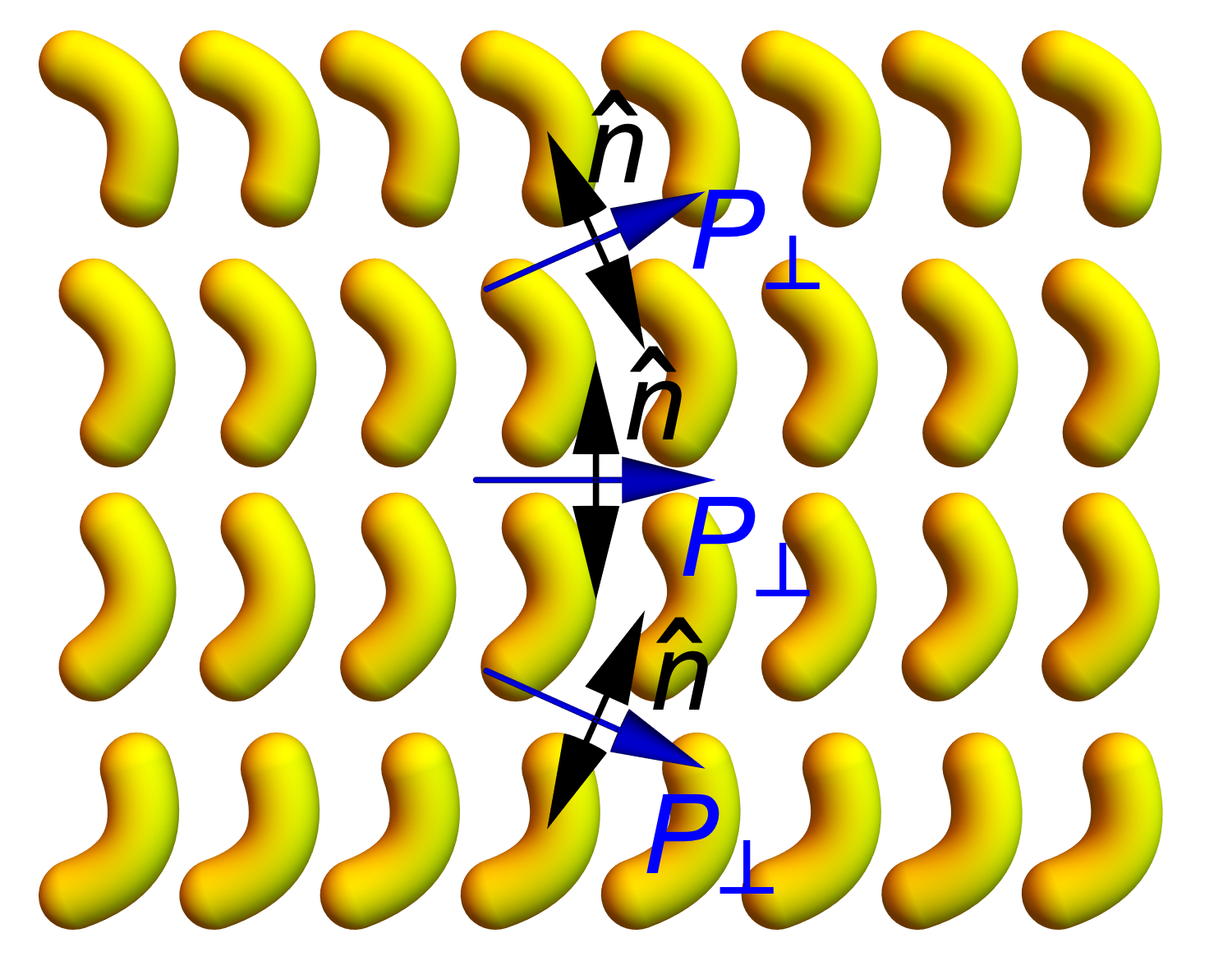}} &
\resizebox{!}{2.6cm}{\includegraphics{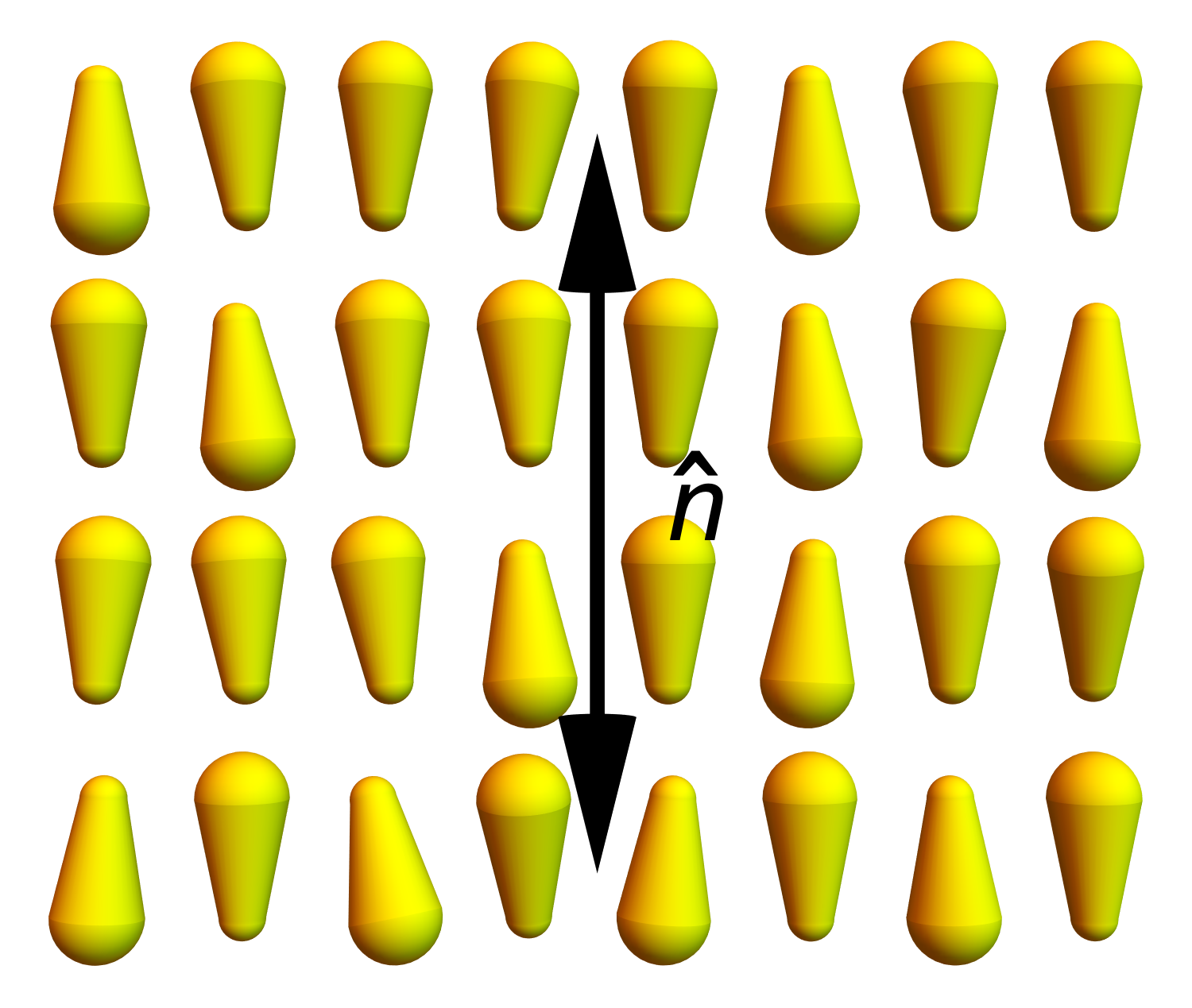}} &
\resizebox{!}{2.6cm}{\includegraphics{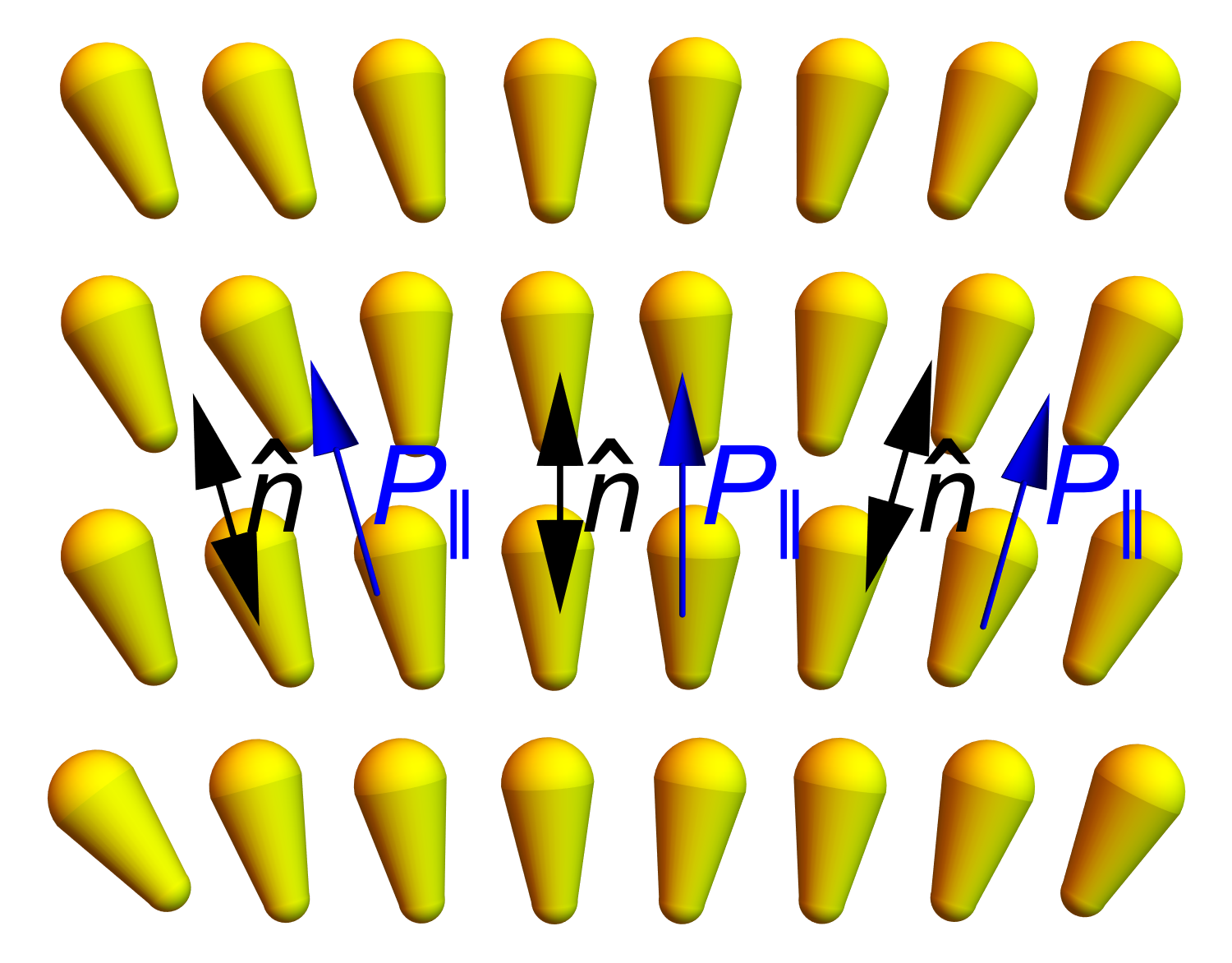}} \\
\end{tabular}
\caption{Schematic illustration of (a,b)~the bend flexoelectric effect in bent-core liquid crystals, and (c,d)~the splay flexoelectric effect in pear-shaped liquid crystals.  Figure adapted with permission from Reference~\cite{Jakli2018}, copyright 2018 American Physical Society.  Based on the concept of Reference~\cite{Meyer1969}.}
\label{fig:flexoelectric}
\end{figure}

\subsubsection{Bend}  Because bend is a vector perpendicular to $\hat{\bm{n}}$, it induces molecular order that is also a vector perpendicular to $\hat{\bm{n}}$.  Physically, this effect can be understood through the mechanism shown in Figures~\ref{fig:flexoelectric}(a,b).  Suppose the molecules are shaped like bananas, as in bent-core liquid crystals.  In the undeformed nematic phase, the long axes of the bananas are aligned along $\hat{\bm{n}}$, but the transverse orientations are random.  However, if a bend is applied, it breaks the symmetry of rotation about $\hat{\bm{n}}$.  Hence, the transverse orientations tend to align, and the system acquires polar order.  This statistical ordering can be represented as a dimensionless vector $\bm{P}_\perp$.  If the molecules have any internal electric dipole moments, then this order can be observed as a net electrostatic polarization perpendicular to $\hat{\bm{n}}$.

Mathematically, this effect can be modeled through a statistical theory~\cite{Meyer1976,Shamid2013}, by adding two terms to the free energy $F=F_\text{nem}-\lambda_\perp\bm{B}\cdot\bm{P}_\perp+\frac{1}{2}\mu_\perp|\bm{P}_\perp|^2$.  The first extra term is a coupling between the bend vector and polar order, and the second represents the free energy cost of developing polar order.  Because this free energy cost involves entropy, it presumably varies with temperature.  Minimizing the free energy over polar order gives $\bm{P}_\perp=\lambda_\perp\bm{B}/\mu_\perp$, proportional to the imposed bend.

When $\bm{P}_\perp$ relaxes to the favored value, the last two terms contribute $-\lambda_\perp^2 |\bm{B}|^2/(2\mu_\perp)$ to the free energy.  Hence, they change the effective elastic constant for bend from $K_{33}$ to the renormalized $K_{33}^R=K_{33}-\lambda_\perp^2/\mu_\perp$.  This mechanism may account for the anomalously low bend elastic constant observed in bent-core liquid crystals, as discussed in the review~\cite{Jakli2018}. 

\subsubsection{Splay}  Because splay is a vector parallel to $\hat{\bm{n}}$, it induces molecular order that is also a vector parallel to $\hat{\bm{n}}$.  This effect is illustrated in Figures~\ref{fig:flexoelectric}(c,d), which shows molecules shaped like pears.  In the undeformed nematic phase, half of the molecules are oriented upward along $\hat{\bm{n}}$, and half are oriented downward.  However, if a splay is applied, it breaks the symmetry between those two orientations, creating a population difference.  The system acquires polar order along $\hat{\bm{n}}$, which can be represented by a dimensionless vector $\bm{P}_\parallel$.  Again, if the molecules have any internal electric dipole moments, then this order leads to a net electrostatic polarization, now parallel to $\hat{\bm{n}}$.

This effect can be modeled through a statistical theory~\cite{Meyer1976,Dhakal2010}, similar to the bend case.  The free energy must now be $F=F_\text{nem}-\lambda_\parallel\bm{S}\cdot\bm{P}_\parallel+\frac{1}{2}\mu_\parallel|\bm{P}_\parallel|^2$, with both extra terms as before.  Minimizing the free energy over polar order gives $\bm{P}_\parallel=\lambda_\parallel\bm{S}/\mu_\parallel$, proportional to the imposed splay vector.  When $\bm{P}_\parallel$ takes this value, the last two terms give a negative correction of $-\lambda_\parallel^2/\mu_\parallel$ to the effective elastic constant for splay.

In a typical experimental liquid crystal, the arbitrary molecular shape includes some banana-like component and some pear-like component.  Hence, most nematic materials exhibit both bend and splay flexoelectricity.

\subsubsection{Twist}  Because twist is a pseudoscalar, it can induce molecular order that is also a pseudoscalar.  Physically, we can understand this effect because one chiral conformation of a molecule is more compatible with a specific twisted environment than the mirror-image conformation.  The difference in concentrations of the two opposite conformations, $\psi=\rho_R-\rho_S$, becomes a chiral order parameter that can couple with the twist.  Indeed, one experiment has reported an induced chiral concentration difference induced by twist~\cite{Basu2011}.  Other possible chiral order parameters have also been considered theoretically~\cite{Selinger1993}.

Mathematically, we can construct a theory as in the bend and splay cases.  The free energy can be written as $F=F_\text{nem}-\lambda_T T \psi+\frac{1}{2}\mu_T \psi^2$, where the first extra term is the coupling between twist and chiral order, and the second extra term is the free energy cost of chiral order.  For example, the second term might be the entropy of mixing the chiral conformations.  Minimizing the free energy over chiral order gives $\psi=\lambda_T T/\mu_T$, proportional to the imposed twist.  With this value of $\psi$, the last two terms give a negative correction of $-\lambda_T^2/\mu_T$ to the effective elastic constant for twist.  We speculate that this mechanism could explain the anomalously low twist elastic constant observed in lyotropic chromonic liquid crystals~\cite{Davidson2015}.  Those materials are composed of molecular aggregates, which might easily deform into chiral conformations that are compatible with the twist.

Because the argument for twist is analogous to the argument for bend and splay flexoelectricity, we will consider this phenomenon as ``generalized flexoelectricity.''  Of course, this concept is not electrical.  The induced order is a pseudoscalar, not a vector, and hence it cannot be associated with an electric dipole moment.

\subsubsection{$\bm{\Delta}$ mode}  Formally, we can apply the same argument to the $\bm{\Delta}$ mode.  Because the deformation $\Delta_{ij}n_k$ is a third-rank tensor, it couples with an octupolar order parameter $O_{ijk}$, which is also a third-rank tensor.  The free energy can be written as $F=F_\text{nem}-\lambda_\Delta \Delta_{ij}n_k O_{ijk} + \frac{1}{2}\mu_\Delta O_{ijk}O_{ijk}$.  Minimizing this free energy over octupolar order then gives $O_{ijk}=\lambda_\Delta \Delta_{ij}n_k/\mu_\Delta$, proportional to the imposed $\bm{\Delta}$ deformation.  This result leads to a negative renormalization of the elastic constant $K_{24}$ for the $\bm{\Delta}$ mode.

The tensor $O_{ijk}$ is not the most general third-rank tensor, because it must have the same structure as $\Delta_{ij}n_k$, with its first two legs traceless, symmetric, and perpendicular to $\hat{\bm{n}}$, and its third leg parallel to $\hat{\bm{n}}$.  Such a tensor has two degrees of freedom (beyond the two degrees of freedom in $\hat{\bm{n}}$ itself), and can be visualized as the tetrahedron in Figure~\ref{fig:modes}(d).

Tetrahedral order was studied in the context of bent-core liquid crystals by Lubensky and Radzihovsky~\cite{Lubensky2002}, and it was mathematically characterized by Gaeta and Virga~\cite{Gaeta2016}.  In Section 2.6, we will suggest two ways in which this order can be achieved by combining biaxial nematic order with $\bm{P}_\parallel$, or combining biaxial nematic order with chiral order.

\subsection{$N+\epsilon$ phases}

In the previous section, we showed that $\bm{P}_\perp$, $\bm{P}_\parallel$, chiral, and octupolar order can be induced by director deformations.  Now, we note that each type of order can form spontaneously.  The system would then have strong nematic order and a small amount of extra order.  We will refer to this possibility as an ``$N+\epsilon$'' phase, with $N$ representing nematic and $\epsilon$ the extra order.  In such a phase, the extra order induces the corresponding director deformation.

For the bend and splay cases, the possibility of spontaneous $\bm{P}_\perp$ or $\bm{P}_\parallel$ order was pointed out by Meyer in lectures at Les Houches~\cite{Meyer1976}, shortly after his discovery of flexoelectricity.  His theory went almost unnoticed, perhaps because there were no relevant experiments at the time, and the same concept was developed again much later~\cite{Shamid2013,Dhakal2010}.  Here, we review the concept and apply it also to twist and the $\bm{\Delta}$ mode.

\subsubsection{Bend}  In section 2.2.1, we developed a quadratic free energy for coupled bend and polar order.  This free energy is positive-definite, with a minimum at $\bm{B}=0$ and $\bm{P}_\perp=0$, provided that $\lambda_\perp^2<K_{33}\mu_\perp$.  However, the free energy becomes unstable if that condition is violated.  The instability may occur if $\mu_\perp$ becomes smaller (perhaps as temperature decreases), or if $K_{33}$ becomes smaller, or if $\lambda_\perp$ becomes larger.  In any of these situations, the state with $\bm{B}=0$ and $\bm{P}_\perp=0$ is no longer a minimum.  The free energy must then be stabilized by some higher-order term, either a higher power (like $\frac{1}{4}\nu_\perp \bm{P}_\perp^4$) or a higher derivative (like $\frac{1}{2}\kappa_\perp(\bm{\nabla}\bm{P}_\perp)^2$).  Competition between the unstable quadratic terms and the stabilizing higher-order term gives a favored polar order, which we can call $\overline{\bm{P}}_\perp$.

As an alternative scenario, the instability might be preempted by a first-order transition.  In that case, the system would go directly from zero polar order to a finite nonzero $\overline{\bm{P}}_\perp$.

Either way, through an instability or a first-order transition, the spontaneous polar order $\overline{\bm{P}}_\perp$ induces a favored bend.  The free energy becomes $F=F_\text{nem}-\lambda_\perp\bm{B}\cdot\overline{\bm{P}}_\perp+$ terms independent of director deformations.  By completing the square and omitting the unimportant constant, we obtain
\begin{equation}
F=\frac{1}{2}(K_{11}-K_{24})S^2 + \frac{1}{2}(K_{22}-K_{24})T^2 + \frac{1}{2}K_{33} |\bm{B}-\overline{\bm{B}}|^2 + K_{24}\Tr(\bm{\Delta}^2).
\label{fbend}
\end{equation}
Hence, the nematic phase with extra $\bm{P}_\perp$ order has a favored bend of $\overline{\bm{B}}=\lambda_\perp\overline{\bm{P}}_\perp/K_{33}$.  In that respect, it is quite different from the conventional nematic phase, which has zero favored bend.

In a highly influential paper~\cite{Dozov2001}, Dozov modeled the instability of a conventional nematic phase to a phase with a favored bend, using a different type of theory.  He did not consider polar order, only the nematic director field, and he found an instability when the bend elastic constant decreased through zero.  We would say that his theory is equivalent to the theory presented here, provided that his bend elastic constant is understood as the renormalized $K_{33}^R=K_{33}-\lambda_\perp^2/\mu_\perp$, which goes to zero when the quadratic part of the free energy becomes unstable.  In our perspective, it does not matter whether the polar order causes the bend or the bend causes the polar order, because these two effects occur together.

\subsubsection{Splay, twist, $\bm{\Delta}$ mode}

The argument for favored bend applies analogously to the other director deformation modes.  If a system has spontaneous polar order $\overline{\bm{P}}_\parallel$, the free energy becomes
\begin{equation}
F=\frac{1}{2}(K_{11}-K_{24})(S-\overline{S})^2 + \frac{1}{2}(K_{22}-K_{24})T^2 + \frac{1}{2}K_{33} |\bm{B}|^2 + K_{24}\Tr(\bm{\Delta}^2),
\label{fsplay}
\end{equation}
with a favored splay of $\overline{S}=\lambda_\parallel\overline{\bm{P}}_\parallel\cdot\hat{\bm{n}}/(K_{11}-K_{24})$.  If a system has spontaneous chiral order $\overline{\psi}$, the free energy becomes
\begin{equation}
F=\frac{1}{2}(K_{11}-K_{24})S^2 + \frac{1}{2}(K_{22}-K_{24})(T-\overline{T})^2 + \frac{1}{2}K_{33} |\bm{B}|^2 + K_{24}\Tr(\bm{\Delta}^2),
\label{ftwist}
\end{equation}
with a favored twist of $\overline{T}=\lambda_T\overline{\psi}/(K_{22}-K_{24})$.  Furthermore, if a system has spontaneous octupolar order $\overline{O}_{ijk}$, the free energy becomes
\begin{equation}
F=\frac{1}{2}(K_{11}-K_{24})S^2 + \frac{1}{2}(K_{22}-K_{24})T^2 + \frac{1}{2}K_{33} |\bm{B}|^2 + K_{24}\Tr[(\bm{\Delta}-\overline{\bm{\Delta}})^2],
\label{fdelta}
\end{equation}
with a favored deformation $\overline{\Delta}_{ij}=\lambda_\Delta\overline{O}_{ijk}n_k/(2K_{24})$.

As an aside, a liquid crystal might have more than one of these four types of molecular order.  In that case, the free energy would include more than one linear term favoring a deformation, as well as cross terms.  For example, a liquid crystal with chiral and parallel polar order would have a cross term of the form $(\overline{\bm{P}}_\parallel\cdot\hat{\bm{n}}S)(\overline{\psi}T)$, known as the $K_{12}$ term.

Of the four possibilities, spontaneous chiral order is by far the most common.  Indeed, chirality is ubiquitous in both natural and synthetic materials.  Molecules often form in some chiral conformation, and they must overcome a huge energy barrier to convert into the mirror-image conformation.  Hence, a material can remain in a chiral structure, characterized by some permanent chiral order parameter.  By comparison, polar order $\bm{P}_\perp$ or $\bm{P}_\parallel$ is much less common, and octupolar order has rarely been studied at all.

Perhaps chirality is the most common type of order because it is a pseudoscalar, and hence is associated with an internal feature of the molecular structure.  By contrast, the other types of order are vectors or tensors, which depend on how the molecules are oriented in space.  We speculate that an internal molecular property is very hard, with high energy barriers to change, while orientational properties are softer, with much lower energy barriers.

Even if the case of favored twist is common, while the other cases are unusual, we can still consider all of these possibilities following the same logical argument.  All four scenarios give ideal local structures with specific nonzero deformations in the director field.

\subsection{Incompatibility of ideal local structures}

Looking at the free energy of Equation~\ref{fbend}, \ref{fsplay}, \ref{ftwist}, or \ref{fdelta}, one might naively think that it is easy to find the ground state:  Just set each of the deformation modes, $S$, $T$, $\bm{B}$, and $\bm{\Delta}$, equal to its favored value.  However, the problem is actually much more complex, because the four modes are not independent of each other.  Rather, all four of the modes must be derived from the same director field $\hat{\bm{n}}(\bm{r})$.  Hence, one must ask:  Is there any director field that has the favored $S$, $T$, $\bm{B}$, and $\bm{\Delta}$, not just at one position, but everywhere in space?

In general, the answer to that question is no.  It is impossible to construct a director field that has only a single deformation mode with constant nonzero magnitude, and all other deformation modes equal to zero.  The technical term for this result is ``incompatibility'':  A single pure deformation mode is not compatible with the Euclidean geometry of 3D space.

This answer has emerged over many years of theoretical research.  In his Les Houches lectures in the 1970s, Meyer~\cite{Meyer1976} already stated that it was impossible to fill continuous 3D space with pure constant bend or splay.  He did not yet have a distinction between pure twist (which is double twist) compared with cholesteric twist.

In the 1980s, Sethna et al.~\cite{Sethna1983} considered the problem of filling space with pure constant double twist.  They noted that this construction is impossible in ordinary 3D Euclidean flat space.  However, as a mathematical generalization, they found that it is possible in the 3D curved non-Euclidean geometry of a hypersphere, known as $S^3$, with a curvature radius of $2/\overline{T}$.  (We emphasize that 3D non-Euclidean geometry means that the space itself is curved, as in general relativity.  It does not mean a volume enclosed by a curved surface.)

More recently, Niv and Efrati~\cite{Niv2018} developed a mathematical formalism to assess what director deformations are compatible with 2D Euclidean or non-Euclidean geometry.  In 2D, the only deformations are bend and splay; there is no twist or $\bm{\Delta}$.  They showed that any uniform deformations must satisfy $|\bm{B}|^2 + S^2 = -K_G$, where $K_G$ is the Gaussian curvature of the surface.  Hence, pure constant bend or splay may occur in a negatively curved surface, but not in a flat or positively curved surface.

Virga~\cite{Virga2019} used the formalism of four director deformation modes to determine what director deformations are compatible with 3D Euclidean geometry.  He showed that certain uniform combinations are possible---either twist and $\bm{\Delta}$, or bend, twist, and $\bm{\Delta}$.  However, it is impossible to construct a director field in this geometry with just one pure constant deformation mode.  Furthermore, Sadoc et al.~\cite{Sadoc2020} generalized the non-Euclidean geometry of Sethna et al.\ to all four deformation modes and all possible homogeneous curved spaces, known as the Thurston geometries.  They found that each pure constant deformation mode is compatible with specific types of curved space, but not with flat Euclidean space.

\subsection{Achievable global phases}

Because it is impossible fill space with a pure constant deformation mode, it is not easy to minimize the free energy of Equation~\ref{fbend}, \ref{fsplay}, \ref{ftwist}, or \ref{fdelta}.  In that situation, a liquid crystal has two alternatives.  First, it can fill space with an allowed combination of deformation modes, including the favored mode and some other mode that costs free energy.  Second, it can break up space into domains of the favored mode, separated by domain walls or defects.  Here, we will see examples of both types of behavior.

\subsubsection{Twist}  Let us begin with a liquid crystal with chiral order, because that is the most common case.  The free energy of Equation~\ref{ftwist} favors a certain nonzero twist $\overline{T}$, along with zero bend, splay, and $\bm{\Delta}$.  However, there is no director field in 3D Euclidean space with that set of deformation modes.  Hence, the liquid crystal experiences geometric frustration.

\begin{figure}
\begin{tabular}{@{}ccc@{}}
(a) Cholesteric & (b) Blue phase 1 & (c) Blue phase 2 \\
\resizebox{3.8cm}{!}{\includegraphics{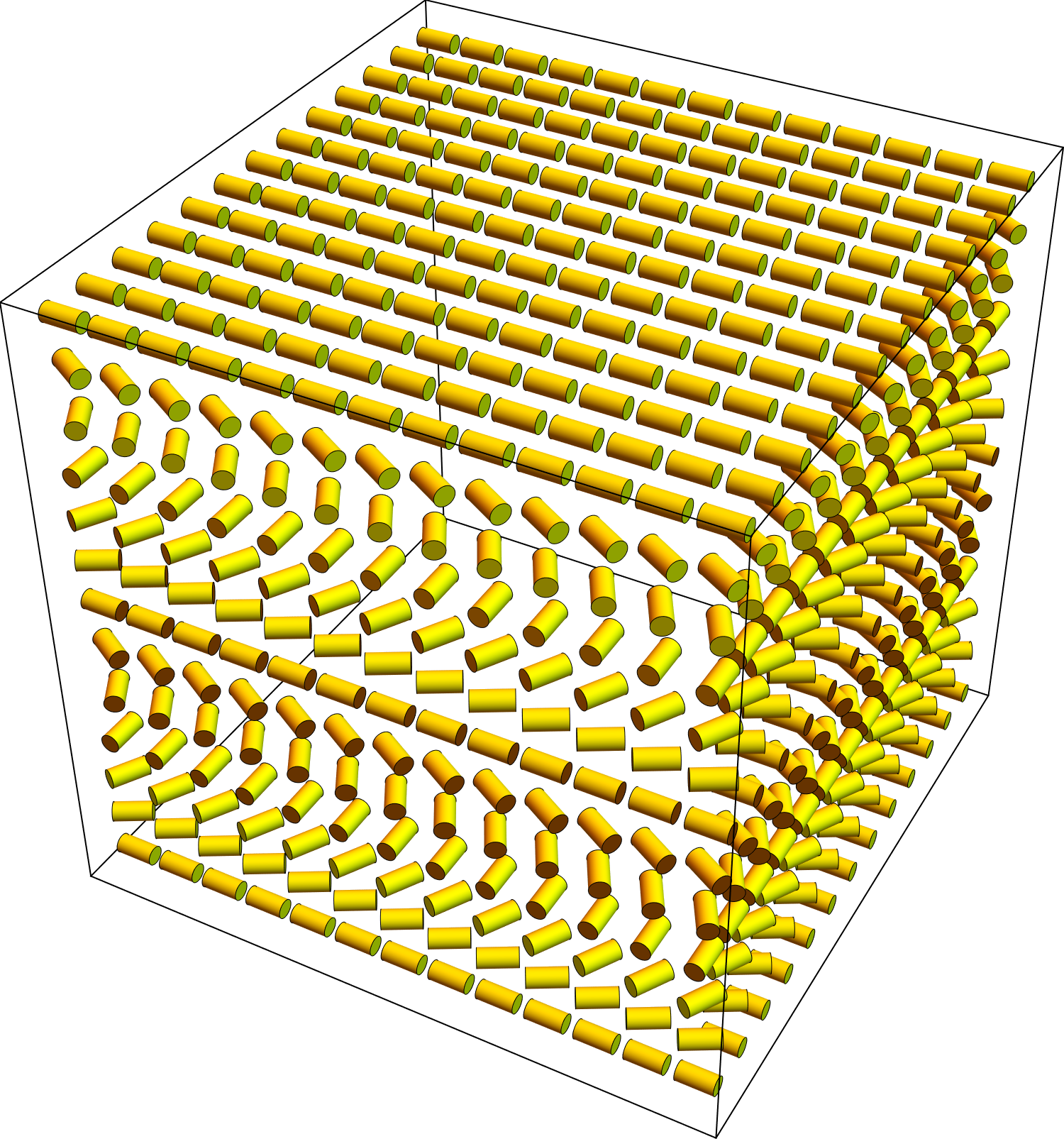}} &
\resizebox{3.8cm}{!}{\includegraphics{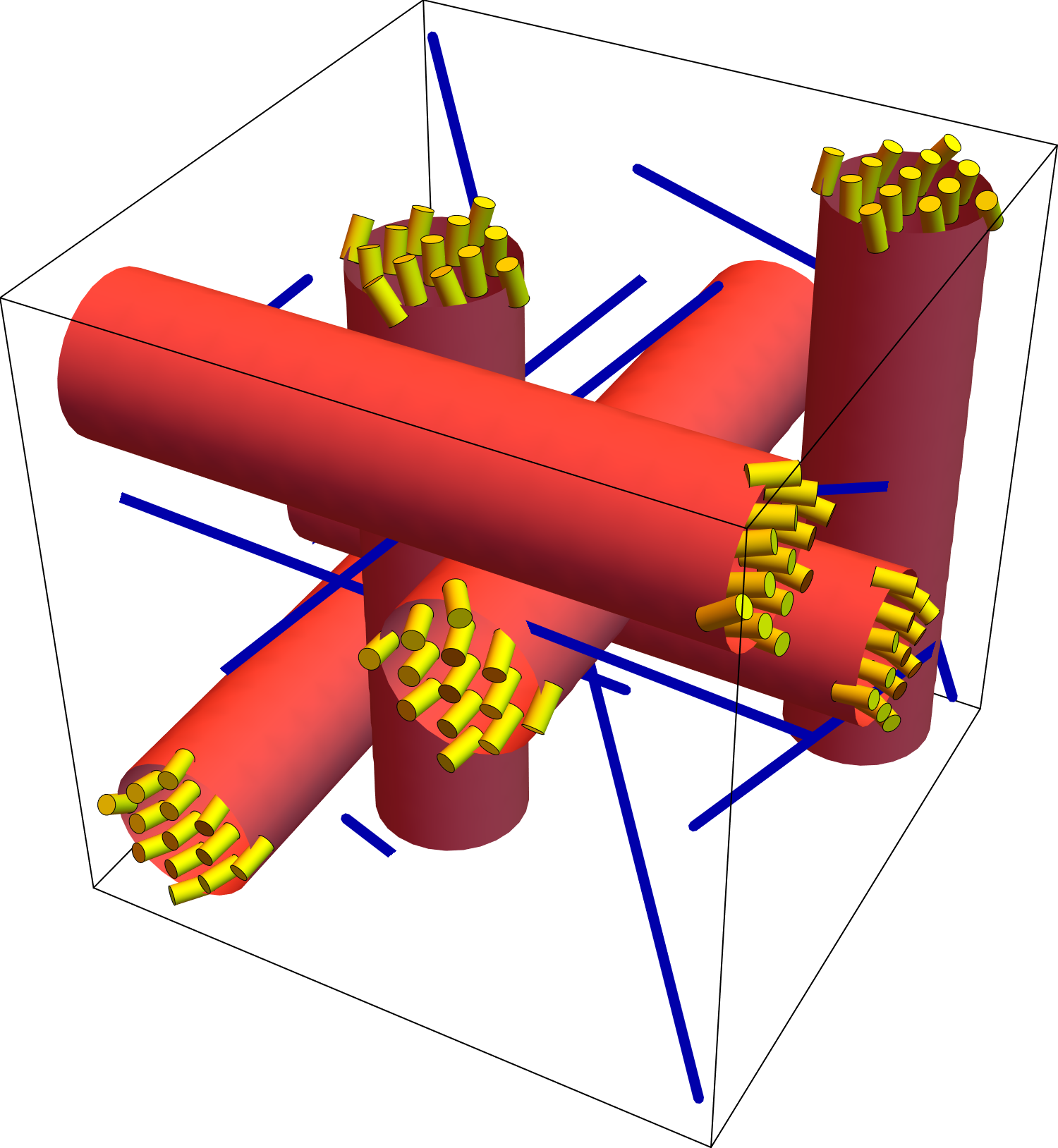}} &
\resizebox{3.8cm}{!}{\includegraphics{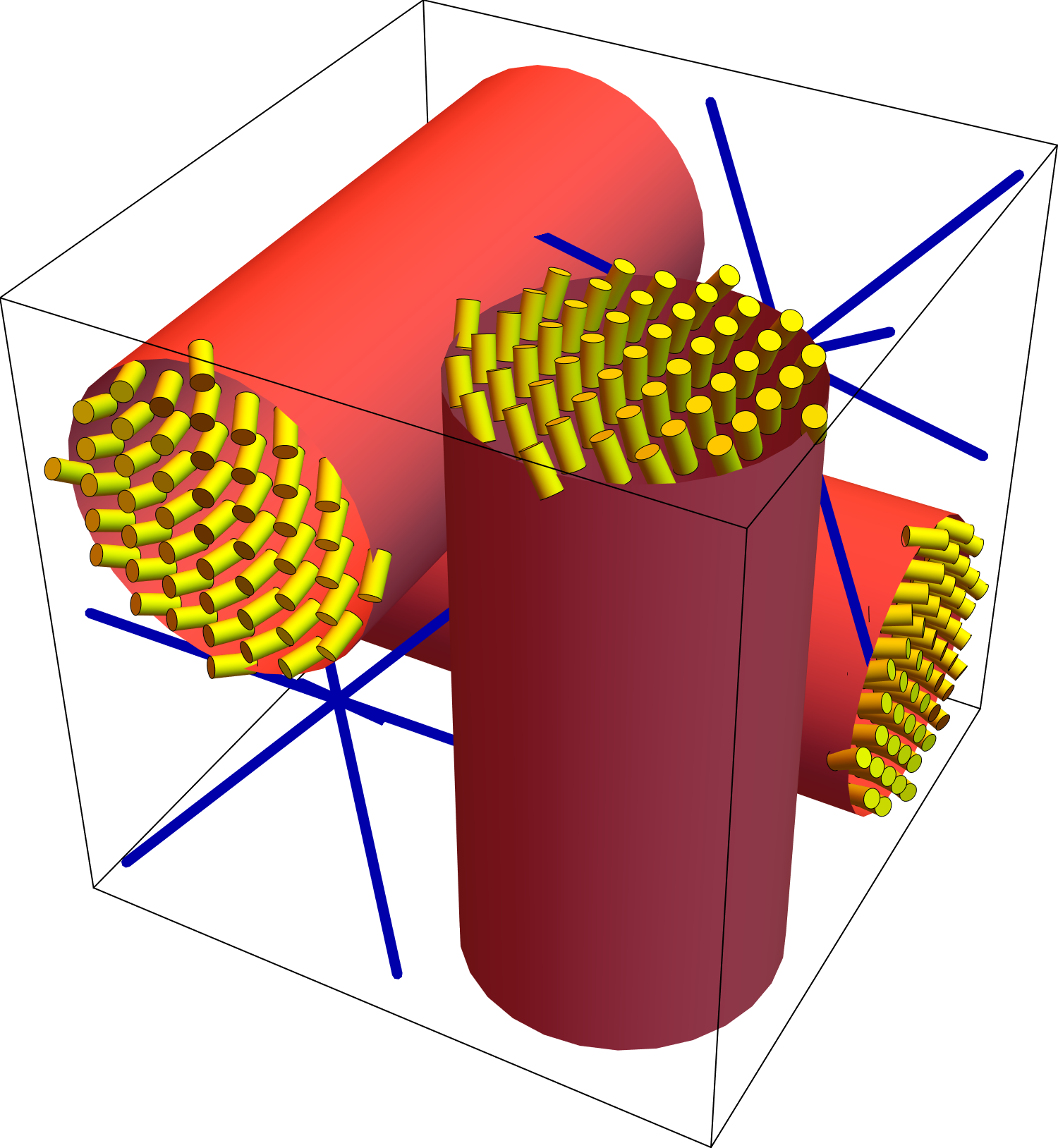}} \\
\end{tabular}
\caption{Three modulated nematic phases induced by chirality:  (a) Cholesteric.  (b) Blue phase 1.  (c) Blue phase 2.  In all cases, the small yellow cylinders represent the director field.  In the blue phases, the large red cylinders represent the double twist tubes, and the blue lines represent the disclinations.  For clarity, the director field is shown only inside the double twist tubes, although it is defined everywhere except along the disclinations.}
\label{fig:chiral}
\end{figure}

\paragraph{Cholesteric phase}  One way to resolve the frustration is to form a cholesteric phase, as shown in Figure~\ref{fig:chiral}(a).  In this phase, the director field forms a helix with a single helical axis.  This structure has twist everywhere, but it is not pure twist, because it also has a component of the $\bm{\Delta}$ mode.  In the cholesteric phase, the liquid crystal must accept some $\bm{\Delta}$ mode, which gives a positive contribution to the free energy, in order to achieve the favored twist, which gives a negative contribution to the free energy.

For an explicit calculation, we choose coordinates so that the helical axis is along $\hat{\bm{z}}$, and write the director as $\hat{\bm{n}}(\bm{r})=(\cos qz,\sin qz,0)$.  The deformation modes are then
\begin{equation}
T=-q,\quad\bm{B}=0,\quad S=0,\quad\bm{\Delta}=\frac{q}{2}
\begin{pmatrix}
0 & 0 & -\sin qz \\
0 & 0 & \cos qz \\
-\sin qz & \cos qz & 0
\end{pmatrix},
\end{equation}
and the free energy becomes $F=\frac{1}{2}(K_{22}-K_{24})(q+\overline{T})^2 + \frac{1}{2}K_{24}q^2$.  Minimizing over the wavevector gives $q=-(K_{22}-K_{24})\overline{T}/K_{22}=-\lambda_T \overline{\psi}/K_{22}$, proportional to the spontaneous chiral order $\overline{\psi}$.  The effects of $K_{24}$ cancel out, and the wavevector depends on $K_{22}$.

\paragraph{Blue phases}  An alternative way to resolve the frustration is to form a blue phase, as shown in Figures~\ref{fig:chiral}(b,c).  Blue phases are complex structures with 3D modulations of the director.  The modulated director field can be regarded as a lattice of double-twist tubes, running in the $x$, $y$, and $z$ directions.  At the center of each double-twist tube, the director deformation is approximately pure twist, which matches the ideal local structure.  However, as we move outward from the center, the twist is mixed with other deformation modes.  Hence, each tube has a certain maximum size, beyond which the deformation is no longer mostly twist.  The space between the tubes is filled with other director deformation modes and also with disclinations, which are singular lines on which the director is undefined.

The two structures in Figures~\ref{fig:chiral}(b,c), known as blue phases 1 and 2, have different arrangements of the double-twist tubes and disclinations, with body-centered cubic and simple cubic periodicity, respectively.  A third alternative, blue phase 3, has a disordered configuration of double-twist tubes and disclinations.  (In these phases, the standard description of the director field in terms of double-twist tubes is not exact, as discussed by Machon and Alexander~\cite{Machon2016}.)

Which has lower free energy: a cholesteric or a blue phase?  Both possibilities have favorable, negative contributions to the free energy arising from the twist.  However, they each make different sacrifices to achieve that twist.  The cholesteric phase must have unfavorable $\bm{\Delta}$ deformations, with a free energy cost proportional to $K_{24}$.  By contrast, blue phases must have disclinations, which have a certain line energy per length.  Hence, as a rough theory, we can say that the cholesteric phase is preferred if $K_{24}$ is small compared with the disclination line energy, while blue phases are preferred if the line energy is small compared with $K_{24}$~\cite{Selinger2018}.  More quantitatively detailed theories have been developed~\cite{Wright1989}, but we believe that this simple argument captures the essential difference between the phases.

We recognize that the discussion in this section is not the usual way of describing the cholesteric phase.  Researchers often neglect the $\bm{\Delta}$ mode, and consider the cholesteric as if it had pure twist, with no frustration.  However, we suggest that this discussion in terms of the $\bm{\Delta}$ mode provides a clearer understanding of cholesteric structure, and especially of the free energy balance between cholesteric and blue phases.  For biaxiality in the cholesteric phase, see Section 2.6.

\subsubsection{Bend}  Now consider a liquid crystal with polar order $\bm{P}_\perp$, perpendicular to the local director.  The free energy of Equation~\ref{fbend} favors a bend $\overline{\bm{B}}$ with a specific nonzero magnitude, along with zero splay, twist, and $\bm{\Delta}$.  As discussed in Section 2.4, no director field in 3D Euclidean space can have that set of deformations.  Thus, this system provides another example of geometric frustration.

\begin{figure}
\begin{tabular}{@{}ccc@{}}
(a) Twist-bend nematic & (b) Splay-bend nematic & (c) Hexagonal \\
\rule[-1em]{0pt}{1em}%
\resizebox{!}{3.8cm}{\includegraphics{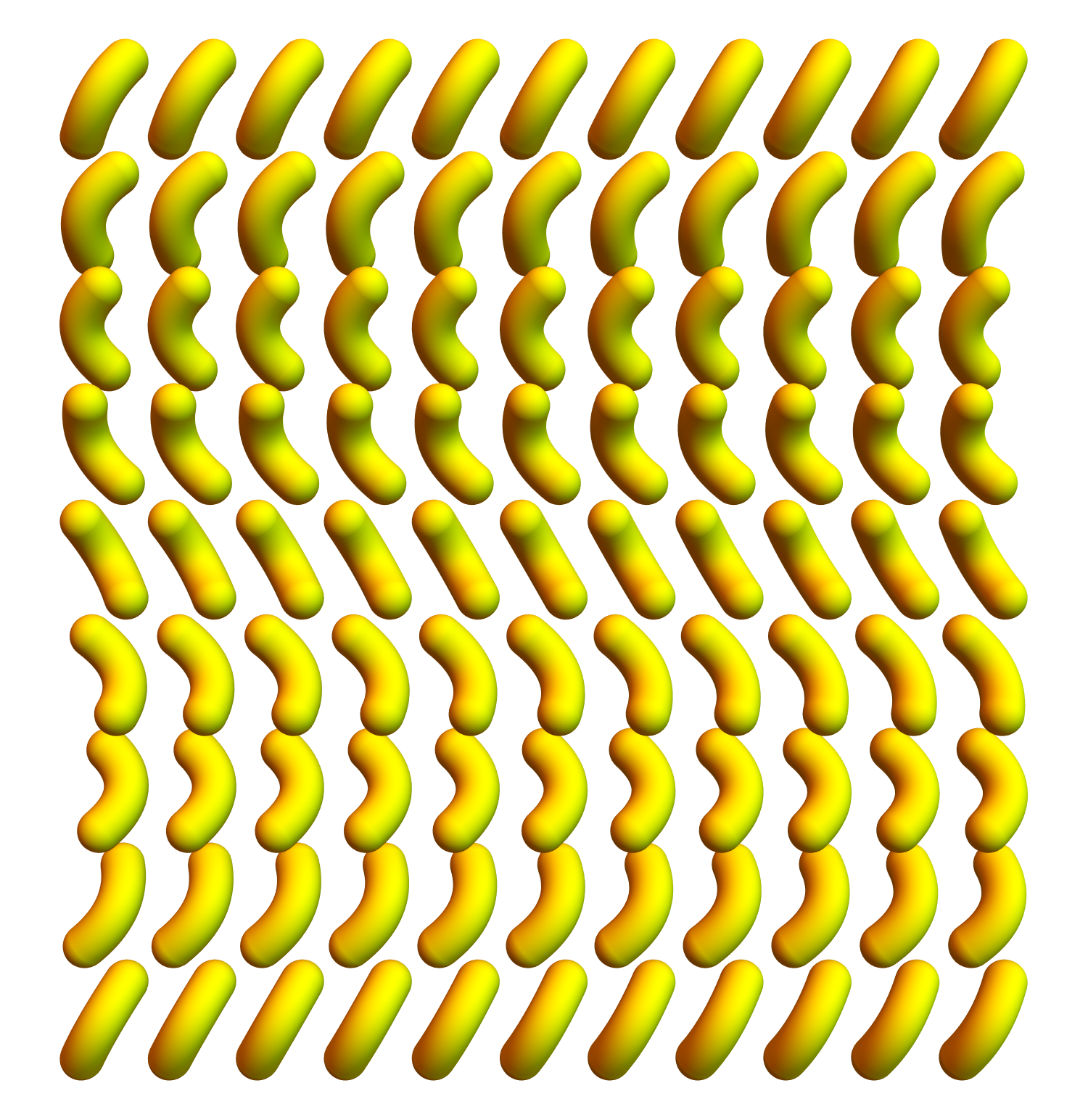}} &
\resizebox{!}{3.8cm}{\includegraphics{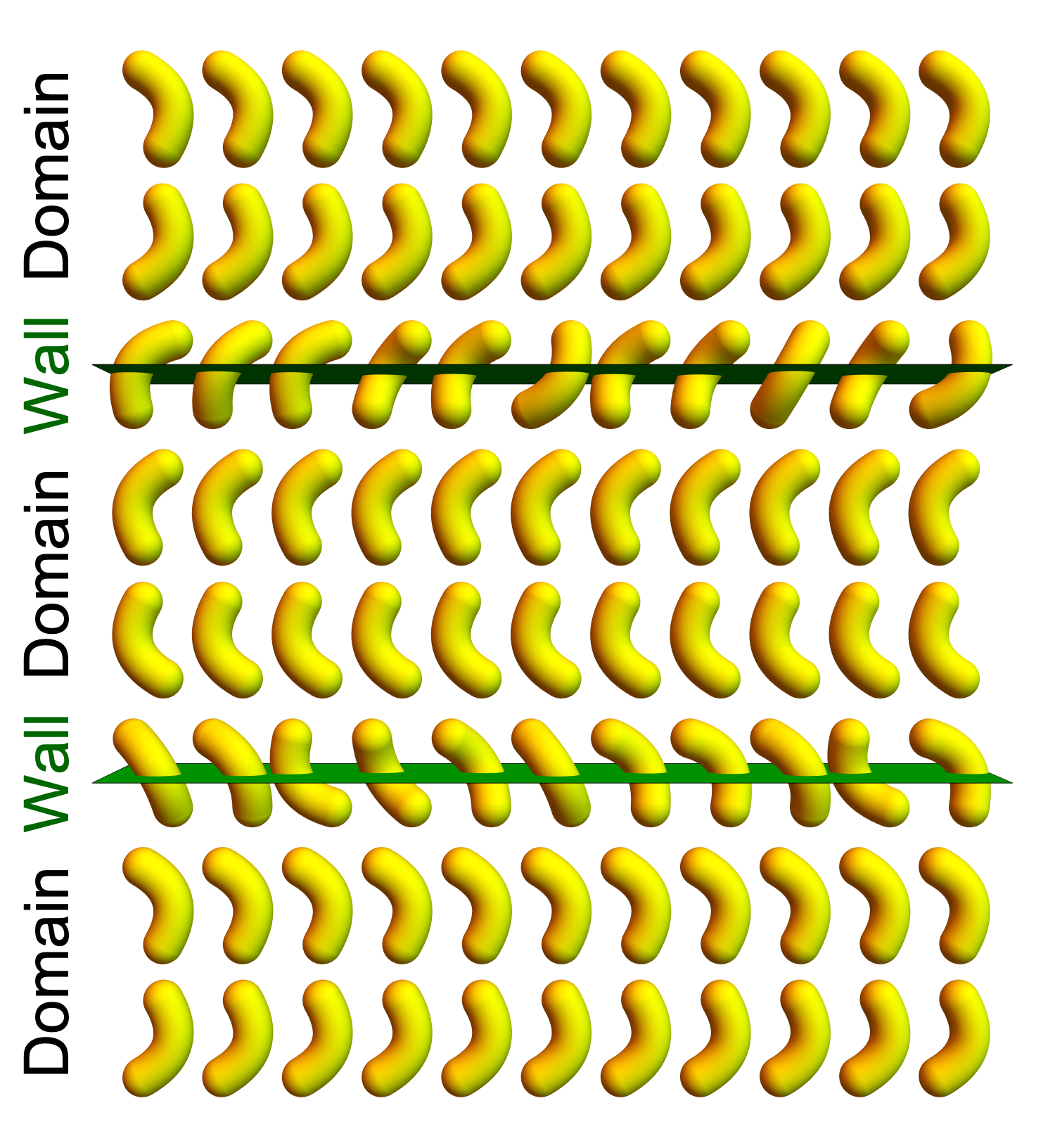}} &
\resizebox{!}{3.8cm}{\includegraphics{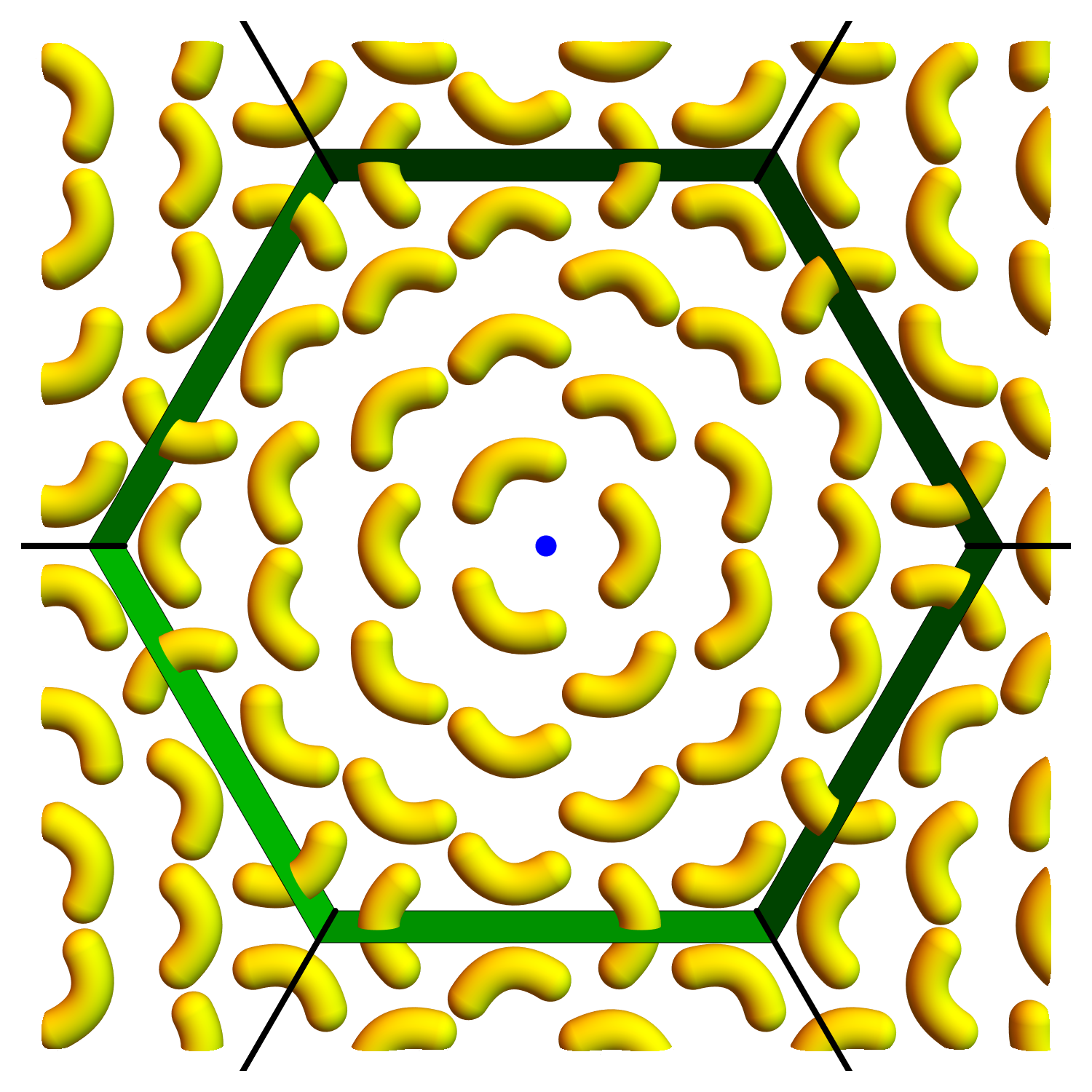}} \\
\end{tabular}
\caption{Three modulated nematic phases induced by perpendicular polar order:  (a)~Twist-bend nematic $N_{TB}$.  (b)~Splay-bend nematic $N_{SB}$.  (c) Hexagonal.  The long axis of the yellow symbols represents the director $\hat{\bm{n}}$, and the transverse orientation represents $\bm{P}_\perp$.  In all three cases, the configuration is extended uniformly in the direction perpendicular to the page.  In the $N_{SB}$ and hexagonal phases, the green planes indicate walls between neighboring domains.}
\label{fig:bend}
\end{figure}

\paragraph{Twist-bend nematic phase}  In response to this frustration, a liquid crystal with $\bm{P}_\perp$ order may form a ``twist-bend nematic'' ($N_{TB}$) phase, as shown in Figure~\ref{fig:bend}(a).  This phase was first predicted by Meyer~\cite{Meyer1976} based on the concept of coupled bend and polarization, and further by Dozov~\cite{Dozov2001} based on the concept of a negative bend elastic constant.  In the $N_{TB}$ phase, the director field is modulated in a helix, which is randomly right- or left-handed.  It maintains a constant cone angle with respect to the helical axis, and hence this structure is sometimes called ``heliconical.''  With the helical axis along $\hat{\bm{z}}$, the director can be written as $\hat{\bm{n}}(\bm{r})=(\sin\beta\cos qz,\sin\beta\sin qz,\cos\beta)$, with a fixed cone angle $\beta$, and the corresponding polar order is $\bm{P}_\perp(\bm{r})=P_0(\sin qz,-\cos qz,0)$.  An explicit calculation shows that this structure has nonzero bend, twist, and $\bm{\Delta}$, and zero splay.  The twist and $\bm{\Delta}$ give positive contributions to the free energy, but they are needed to accompany the favored bend, which gives a greater negative contribution to the free energy.

The $N_{TB}$ phase has been observed experimentally in bent dimer liquid crystals~\cite{Cestari2011,Chen2013,Borshch2013}, using a range of techniques including optical and transmission electron microscopy.  The pitch of the observed structure is about 8 nm.  This length is surprisingly small, comparable to the molecular length scale, in contrast with the micron-scale pitch of a cholesteric phase.  The small pitch implies that the polar order parameter $\bm{P}_\perp$ must be substantial.

The small pitch and large polar order have led to a controversy in the recent literature.  Samulski et al.~\cite{Samulski2020} argue that the experimentally observed phase is not an $N_{TB}$ phase because the polar order is so large; instead they call it a polar, twisted $N_{PT}$ phase.  In response, Dozov and Luckhurst~\cite{Dozov2020} argue that $N_{PT}$ has the same symmetry as $N_{TB}$, and hence is within the family of $N_{TB}$ theories.  We respect both sides in this dispute.  Samulski et al.\ have a reasonable point that the concept of polar order as a perturbation on nematic order works well for small polar order (large pitch), and works less well for large polar order (small pitch).  However, there is no specific point at which it stops working.  It works well enough to predict the correct symmetry of the experimental structure.  Hence, we agree with Dozov and Luckhurst that the experimental phase can reasonably be called $N_{TB}$.

\paragraph{Splay-bend nematic phase}  Apart from the $N_{TB}$ phase, another possible response to $\bm{P}_\perp$ order is the ``splay-bend nematic'' ($N_{SB}$) phase, shown in Figure~\ref{fig:bend}(b), which was also predicted by Meyer~\cite{Meyer1976} and by Dozov~\cite{Dozov2001}.  The $N_{SB}$ structure can be regarded as a series of domains separated by walls.  In each domain, the director deformation is predominantly bend, coupled with $\bm{P}_\perp$, which is approximately the ideal local structure.  Because the ideal structure cannot fill up space, each domain ends with a wall, where $\bm{P}_\perp$ changes sign, so that the bend can go in the opposite direction in the next domain.  The walls can be regarded as defects in the polar order.  They are not as severe as the disclinations in a blue phase, which are defects in the nematic order.

The free energy balance between $N_{TB}$ and $N_{SB}$ depends on the free energy for twist in the $N_{TB}$ phase compared with the free energy for walls in the $N_{SB}$ phase.  For weak polar order (large pitch), the wall free energy is mainly associated with director splay, and hence $N_{TB}$ is favored when $K_{22}$ is small compared with $K_{11}$, as shown by Dozov~\cite{Dozov2001}.  For strong polar order (small pitch), we speculate that the wall free energy should be mainly associated with reducing $\bm{P}_\perp$ away from its ideal value.

\paragraph{Other phases}  The $N_{TB}$ and $N_{SB}$ phases are not the only possibilities.  Other phases might have more complex structures with 2D or 3D modulations in the director and the polar order.  Such possibilities have been considered by Lorman and Mettout~\cite{Lorman1999,Lorman2004}, who regarded them as combinations of vector waves, and by Shamid et al.~\cite{Shamid2014}, who called them ``polar blue phases.''  For example, consider the structure in Figure~\ref{fig:bend}(c).  It consists of a hexagonal lattice of domains, which have bend deformation coupled with $\bm{P}_\perp$.  The domains are separated by walls, similar to the walls in the $N_{SB}$ phase, which are defects in the polar order.  This phase also has disclinations in the director field, at the center of each hexagon and at each vertex where three hexagons meet, which are defects in the nematic order.  Hence, this phase can only form if the free energy cost of the disclination lines is not too large compared with the free energy benefit from the domains.  Similar modulated phases have been seen in experiments on supramolecular liquid crystals by Ungar et al.~\cite{Ungar2003}.

\subsubsection{Splay}  Next, suppose that a liquid crystal has polar order $\bm{P}_\parallel$, parallel to the local director.  The free energy of Equation~\ref{fsplay} then favors a splay $\overline{S}$ of fixed nonzero magnitude, with zero twist, bend, and $\bm{\Delta}$.  As in the previous cases, there is no director field in 3D Euclidean space with pure constant splay.  In fact, there is not even a uniform combination of deformation modes that involves splay.  Hence, the only way for a liquid crystal to accommodate the favored splay is to break up space into splay domains, separated by walls.

\begin{figure}
\begin{tabular}{@{}ccc@{}}
(a) Single splay & (b) Double splay & (c) BCC lattice \\
\resizebox{!}{3.8cm}{\includegraphics{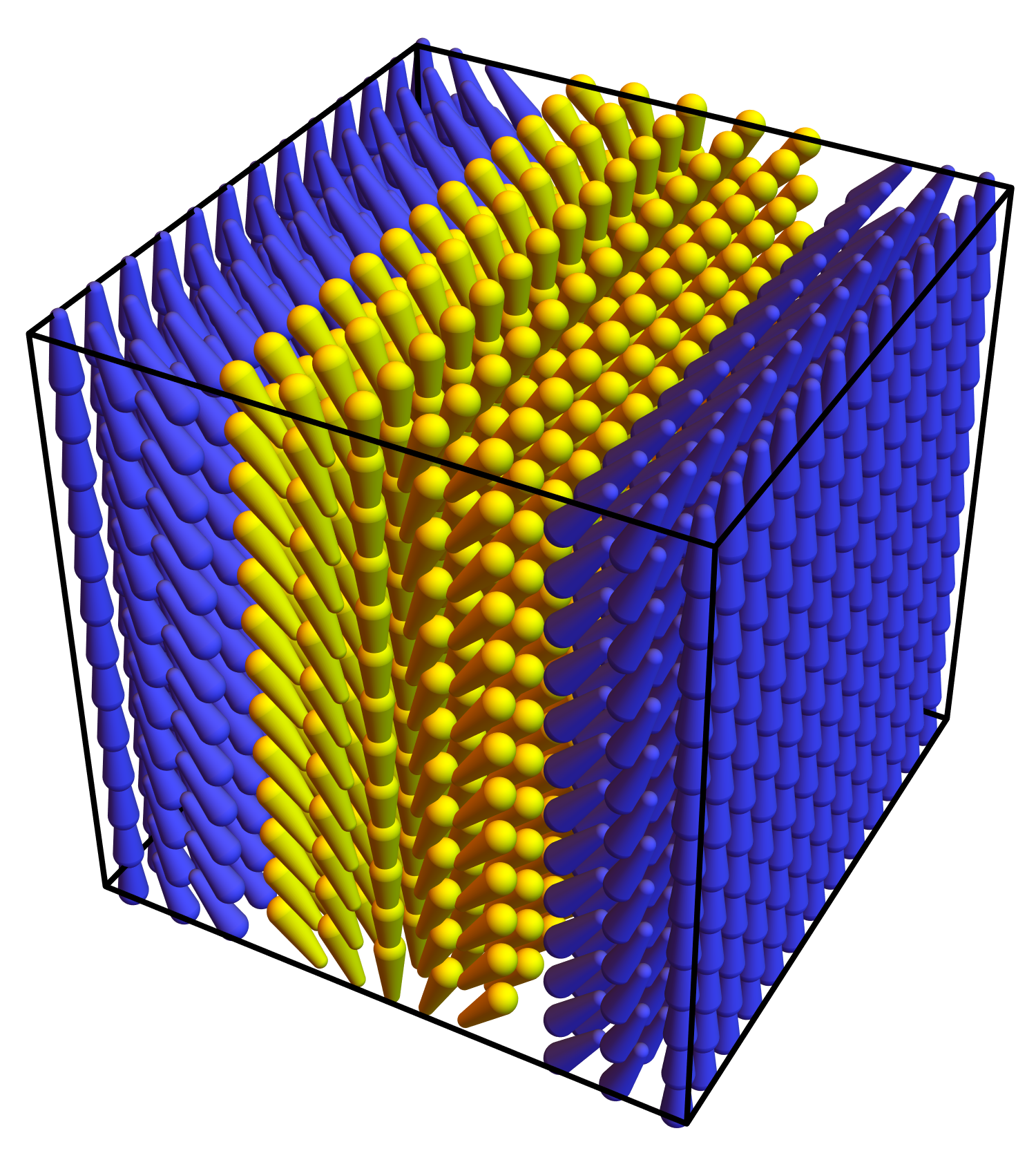}} &
\resizebox{!}{3.8cm}{\includegraphics{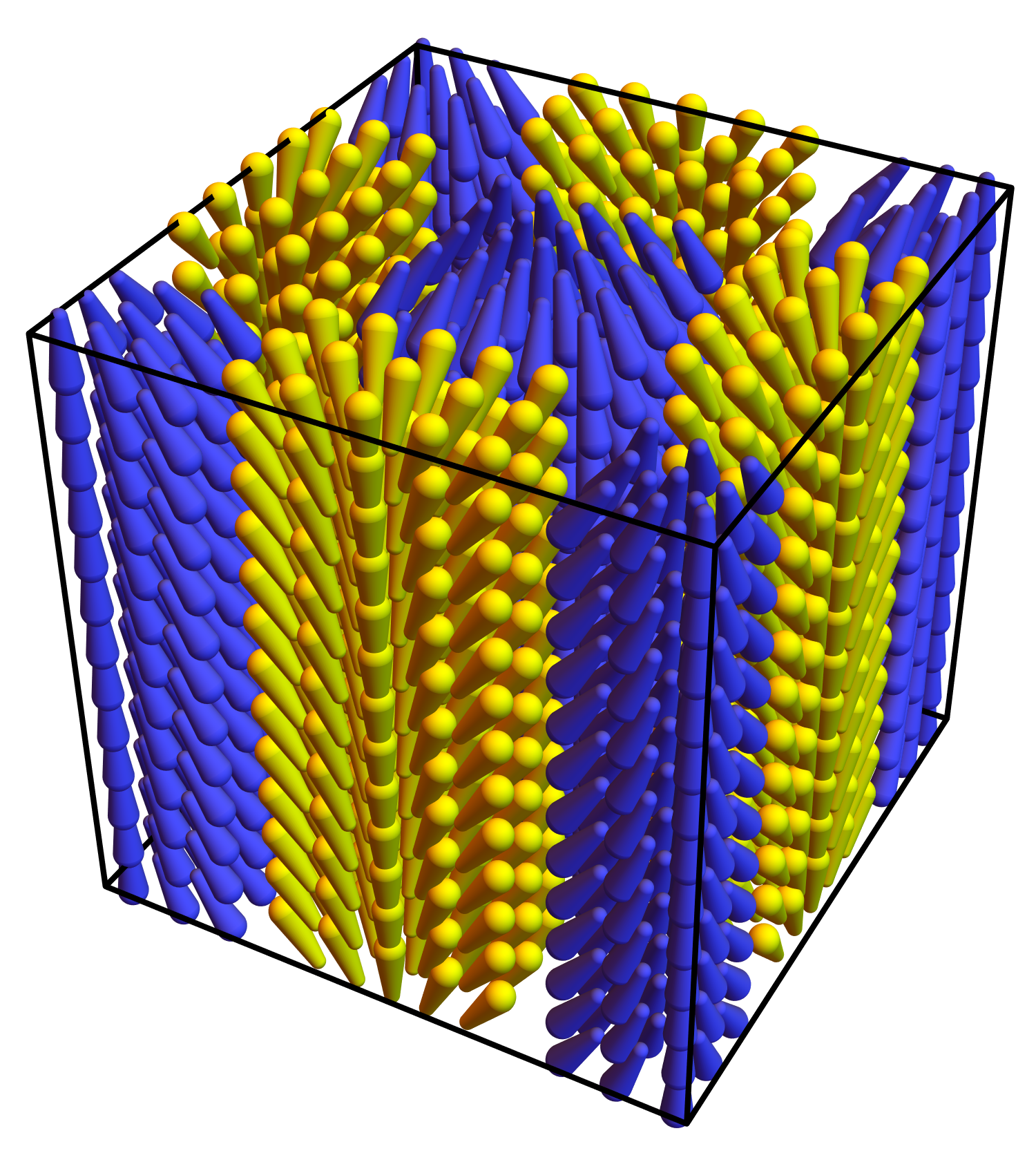}} &
\resizebox{!}{3.8cm}{\includegraphics{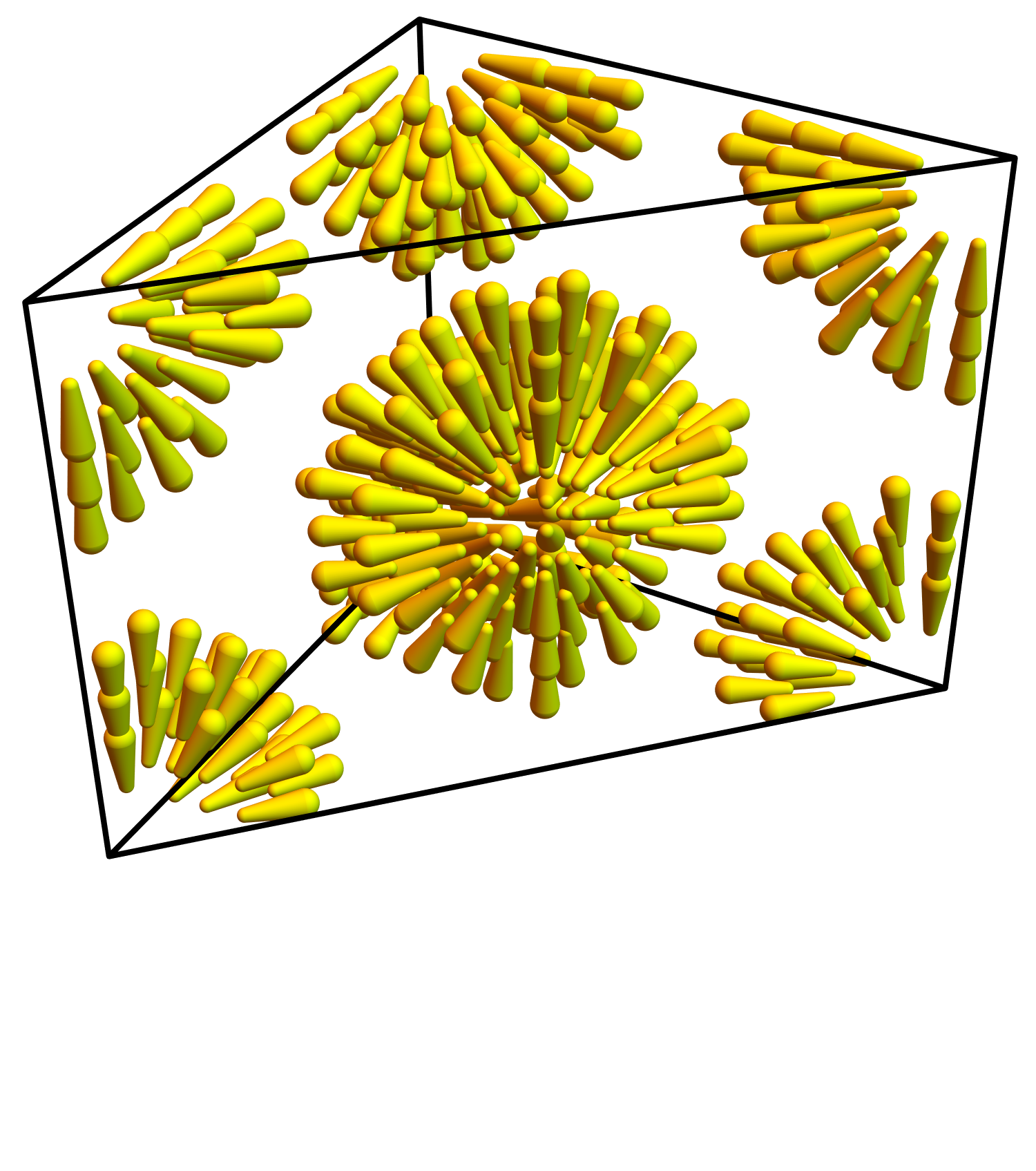}} \\
\end{tabular}
\caption{Three modulated nematic phases induced by parallel polar order:  (a)~Single splay.  (b)~Double splay.  (c)~BCC lattice.  The yellow or blue symbols represent $\hat{\bm{n}}$ and $\bm{P}_\parallel$.  In the BCC lattice, some symbols are omitted for clarity.}
\label{fig:splay}
\end{figure}

\paragraph{Splay nematic phase (1D)}  A phase of splay domains has been reported in experimental work by Mertelj et al.~\cite{Mertelj2018}.  The experimental group called the structure a ``splay nematic'' ($N_S$) phase, and modeled it as in Figure~\ref{fig:splay}(a).  This phase has a 1D modulated structure, which can be regarded as domains separated by walls.  In each domain, the director deformation is single splay, which is a combination of pure double splay and $\bm{\Delta}$.  Each domain ends with a wall in which $\bm{P}_\parallel$ changes sign, so that the splay can go in the opposite direction in the next domain.  Hence, the walls are defects in the polar order, but not in the nematic order.  Chaturvedi and Kamien~\cite{Chaturvedi2019} pointed out theoretically that this phase is very similar to the $N_{SB}$ phase, related by a $90^\circ$ rotation of the director field, so that bend transforms to single splay.

\paragraph{Splay nematic phase (2D)}  An alternative version of the $N_S$ phase has been proposed theoretically by Rosseto and Selinger~\cite{Rosseto2020}.  As shown in Figure~\ref{fig:splay}(b), this structure is a 2D checkerboard lattice of square domains.  At the center of each domain, the director field has pure splay (i.e.\ double splay), coupled with $\bm{P}_\parallel$ order.  At each edge where two square domains meet, there is a wall in which $\bm{P}_\parallel$ changes sign.  At each vertex where four square domains meet, the director has a concentration of the $\bm{\Delta}$ mode.  A calculation shows that this 2D version of the $N_S$ phase has a lower free energy than the 1D version in the limit of small $\bm{P}_\parallel$ order, where the pitch is large~\cite{Rosseto2020}.  However, the free energy comparison is more complex when the $\bm{P}_\parallel$ order is larger, so that the pitch is smaller.

\paragraph{Other phases}  As with twist and bend, splay might also induce more complex phases with 3D modulated structures, which can be regarded as combinations of vector waves~\cite{Lorman1999,Lorman2004} or as polar blue phases~\cite{Shamid2014}.  For example, Figure~\ref{fig:splay}(c) shows a body-centered cubic (BCC) lattice of radial hedgehogs in the director field.  Each hedgehog has pure splay (i.e.\ double splay), coupled with $\bm{P}_\parallel$, and the walls between hedgehogs (not shown in the figure) are defects in the polar order.  This structure also has defects in the nematic director field at the center of each hedgehog and between the hedgehogs.  Hence, it can only form if the free energy cost of these director defects is not too great.  As with the bend case, these predicted structures are similar to experiments on supramolecular liquid crystals~\cite{Ungar2003}.

Apart from all the modulated phases, a liquid crystal can also form a uniform polar phase, as seen in the phase diagram of Rosseto and Selinger~\cite{Rosseto2020}.  The uniform polar phase occurs if the $\bm{P}_\parallel$ order is very strong, so that the free energy cost of a wall exceeds the free energy benefit of the splay inside a domain.  Indeed, a uniform polar phase with strong $\bm{P}_\parallel$ order has been reported in recent experiments of Chen et al.~\cite{Chen2020}, using the same material in which Mertelj et al.~\cite{Mertelj2018} reported the $N_S$ phase.

\subsubsection{$\bm{\Delta}$ mode}  Let us consider the $\bm{\Delta}$ mode by analogy with twist, bend, and splay.  If a liquid crystal has octupolar order $O_{ijk}$, then the free energy of Equation~\ref{fdelta} favors a deformation $\overline{\Delta}_{ij}$ with a specific nonzero magnitude, with zero twist, bend, and splay.  Once again, this set of deformations cannot be achieved in 3D Euclidean space.  What will the liquid crystal do in response to that geometric frustration?  This question has not yet been studied either theoretically or experimentally, but we can make a few speculations.

One possibility is a phase with a uniform combination of $\bm{\Delta}$ and other modes everywhere in space.  The liquid crystal could form a cholesteric phase as in Figure~\ref{fig:chiral}(a), with a combination of $\bm{\Delta}$ and twist, or an $N_{TB}$ phase as in Figure~\ref{fig:bend}(a), with a combination of $\bm{\Delta}$, twist, and bend.  In either case, the helix would be randomly right- or left-handed, because nothing favors either handedness.

Another possibility is a phase with domains of the ideal local $\bm{\Delta}$ deformation and octupolar order, separated by walls.  This phase might have the same structure as the 1D $N_S$ phase in Figure~\ref{fig:splay}(a), because the single splay in each domain includes a large component of $\bm{\Delta}$.  Alternatively, it might have the same structure as the 2D $N_S$ phase in Figure~\ref{fig:splay}(b), because that phase has concentrations of $\bm{\Delta}$ at each vertex between yellow and blue squares.  In this case, because the modulation is now induced by $\bm{\Delta}$, we would label the region around each vertex as a domain, and the regions between the vertices as the walls.  This subject remains an area for future research.

\subsection{Comments on biaxial nematic order}

So far, we have discussed $N+\epsilon$ phases with four types of extra order---chiral, $\bm{P}_\perp$, $\bm{P}_\parallel$, and octupolar---but we have not yet mentioned biaxial nematic order.  We should consider how it fits into the same theoretical framework.

Biaxial nematic order is another type of extra order, which can occur in addition to the primary nematic order along $\hat{\bm{n}}$.  It occurs if a system develops further alignment along a secondary director in the plane perpendicular to $\hat{\bm{n}}$.  It can be characterized by a symmetric, traceless tensor $\beta_{ij}=b (l_i l_j - m_i m_j)$, where $b$ represents the magnitude of the order, and $\hat{\bm{l}}$ and $\hat{\bm{m}}$ are the two principal axes, perpendicular to each other and to $\hat{\bm{n}}$.

The other four types of extra order have simple bilinear couplings with director deformations of the forms $T\psi$, $\bm{B}\cdot\bm{P}_\perp$, $S\hat{\bm{n}}\cdot\bm{P}_\parallel$, or $\Delta_{ij}n_k O_{ijk}$.  By contrast, biaxial nematic order does not have any such coupling; there is no director deformation with the correct symmetry to couple with it.  One might initially think there could be a coupling of the form $\Delta_{ij}\beta_{ij}$, but $\Delta_{ij}$ is odd in $\hat{\bm{n}}$ while $\beta_{ij}$ is independent of $\hat{\bm{n}}$, so that coupling is forbidden by the symmetry of $\hat{\bm{n}}\leftrightarrow-\hat{\bm{n}}$.  (This is why we now suggest that $\Delta_{ij}$ should be called ``tetrahedral splay'' rather than ``biaxial splay.'')  Biaxial nematic order can couple with bend through $B_i B_j \beta_{ij}$, but that is higher-order in gradients of $\hat{\bm{n}}$.  Hence, $\beta_{ij}$ does not induce any favored director deformation; rather, a biaxial nematic phase tends to be uniform.  In that respect, biaxial nematic order is quite different from the other types of extra order.

Although the $\bm{\Delta}$ deformation does not couple with biaxial nematic order by itself, it can couple with two combinations of biaxial nematic and other order.  First, consider the combination of biaxial nematic and parallel polar order.  With that combination, we can construct the coupling $\Delta_{ij}n_k \beta_{ij}(P_\parallel)_k$, which is allowed by symmetry.  In other words, the combination of $\beta_{ij}$ and $\bm{P}_\parallel$ generates octupolar order, which couples with the $\bm{\Delta}$ deformation.

Second, consider the combination of biaxial nematic order and chirality.  In that case, we can construct the coupling $\Delta_{ij}n_k \beta_{il}\psi\epsilon_{ljk}$, which is again allowed by symmetry.  Hence, the combination of $\beta_{ij}$ and chirality generates octupolar order, which couples with the $\bm{\Delta}$ deformation.  There is a long literature on the presence of biaxiality in cholesteric liquid crystals~\cite{Priest1974,Harris1999,Dhakal2011}.  Here, we suggest that this biaxiality can be understood as a result of the $\bm{\Delta}$ component of cholesteric twist, together with chirality.  It is not directly associated with the pure twist (i.e.\ double twist) component of cholesteric structure.

\section{SMECTIC LIQUID CRYSTALS AND OTHER LAYERED STRUCTURES}

In a smectic liquid crystal, the molecules have nematic orientational order, and they also form a layered structure.  The nematic director field then has a preferred orientation with respect to the layers:  In a smectic-A phase, the director is aligned along the layer normal; in a smectic-C phase, the director has a specific, temperature-dependent tilt with respect to the layer normal.  Apart from bulk smectic phases, other molecular systems self-assemble into structures with a single layer, or with multiple layers.  These self-assembled structures are not necessarily called smectic liquid crystals, but we will consider them together with smectic phases in this section, because the theoretical considerations are similar.

For all of these systems, our goals are to find the ideal local structure, to assess whether this ideal local structure can fill up space, and if not, to determine the best achievable global phase.  We first discuss systems with fluid layers, and then consider the additional constraints that occur if the layers are solid.

\subsection{Constraints on director deformations due to layers}

In a smectic liquid crystal, there is still a director field, and there is still an elastic free energy associated with deformations in the director field.  However, the layer structure imposes important constraints on the director deformations.

To see these constraints, we make two key assumptions.  First, suppose we have a smectic-A phase, with the director is aligned along the local layer normal.  Second, suppose the layers are smooth, continuous 2D objects, which can be characterized by a smooth displacement function.  In that case, each smectic layer $l$ is located at the height $z_l=ld+u(\bm{r})$, where $d$ is the equilibrium layer spacing and $u(\bm{r})$ is the small local displacement.  (The following conclusions apply also for large displacements, but the calculations are simplest for small displacements.)  To lowest order in $u$, the director field is $\hat{\bm{n}}=\pm(\hat{\bm{z}}-\nabla_\perp u)$.  Hence, the director deformations become~\cite{Selinger2018}
\begin{equation}
T=0,\quad
\bm{B}=\nabla_\perp(\partial_z u),\quad
S=\mp\nabla_\perp^2 u,\quad
\bm{\Delta}=\pm
\begin{pmatrix}
\frac{1}{2}(\partial_y^2 u-\partial_x^2 u) & -\partial_x \partial_y u & 0 \\
-\partial_x \partial_y u & \frac{1}{2}(\partial_x^2 u-\partial_y^2 u) & 0 \\
0 & 0 & 0\\
\end{pmatrix}
.
\end{equation}

To interpret these results, let us begin with the twist.  The twist is necessarily zero, just because of our two assumptions above.  It is impossible to have twist, if the director is constrained to be normal to the local layers and the layers are smooth 2D objects.  Next, consider the bend.  At lowest order in $u$, the quantity $\partial_z u$ is variation in layer spacing, compared with the equilibrium:  $\partial_z u > 0$ means that the layers are farther apart, and $\partial_z u < 0$ means that they are closer together.  In a multilayer smectic phase, any variations in this spacing cost free energy, and hence $\partial_z u$ is constrained to be zero on long length scales.  For that reason, the bend is also zero.

The two allowed director deformations are splay and $\bm{\Delta}$.  Both of these deformations are related to the curvature of the smectic layers~\cite{Selinger2018}.  Suppose the curvature tensor has eigenvalues of $\kappa_1$ and $\kappa_2$, the two principal curvatures.  The splay becomes $S={\mp(\kappa_1+\kappa_2)}=2H$, which is twice the mean curvature $H$.  Furthermore, $\bm{\Delta}$ is the traceless part of the curvature tensor, with eigenvalues of $\pm\frac{1}{2}(\kappa_1 -\kappa_2)$ and $0$.  Hence, the standard expression for the saddle-splay becomes $\hat{\bm{n}}(\bm{\nabla}\cdot\hat{\bm{n}})+\hat{\bm{n}}\times(\bm{\nabla}\times\hat{\bm{n}}) = \frac{1}{2}S^2 + \frac{1}{2}T^2 - \Tr(\bm{\Delta}^2) = 2\kappa_1 \kappa_2 =2 K_G$, which is twice the Gaussian curvature $K_G$.  The nematic free energy of Equations~\ref{Fnemconventional}--\ref{Fnem} can then be written as
\begin{equation}
F_\text{nem}=\frac{1}{2}(K_{11}-K_{24})(\kappa_1+\kappa_2)^2 + \frac{1}{2}K_{24}(\kappa_1-\kappa_2)^2
=\frac{1}{2}K_{11}(2H)^2 - 2 K_{24} K_G.
\end{equation}
This expression can be recognized as the Helfrich free energy for membrane curvature~\cite{Helfrich1973}, which is well-known in biophysics.

\subsection{Achievable global phases}

Suppose that a smectic liquid crystal has any of the four favored deformation modes, discussed in the previous section about nematics.  Let us consider what is the ideal local structure, and what is the best achievable global phase, given the layer constraints.  

\subsubsection{Twist}  We begin with a liquid crystal with chiral order, because that is the most common situation.  The nematic free energy of Equation~\ref{ftwist} favors a certain nonzero twist $\overline{T}$.  Furthermore, in a smectic-A phase, the smectic part of the free energy favors a structure of smooth, continuous 2D layers, with the director aligned along the layer normal.  However, it is impossible to construct a director field and layer structure that satisfies all of these requirements.  Thus, the liquid crystal experiences a new form of geometric frustration.

\paragraph{Twist-grain-boundary phases}

\begin{figure}
\tabcolsep-5pt
\begin{tabular}{ccc}
(a) TGB phase & (b) Helical nanofilament & (c) Colloidal membrane \\
\resizebox{4.2cm}{!}{\includegraphics{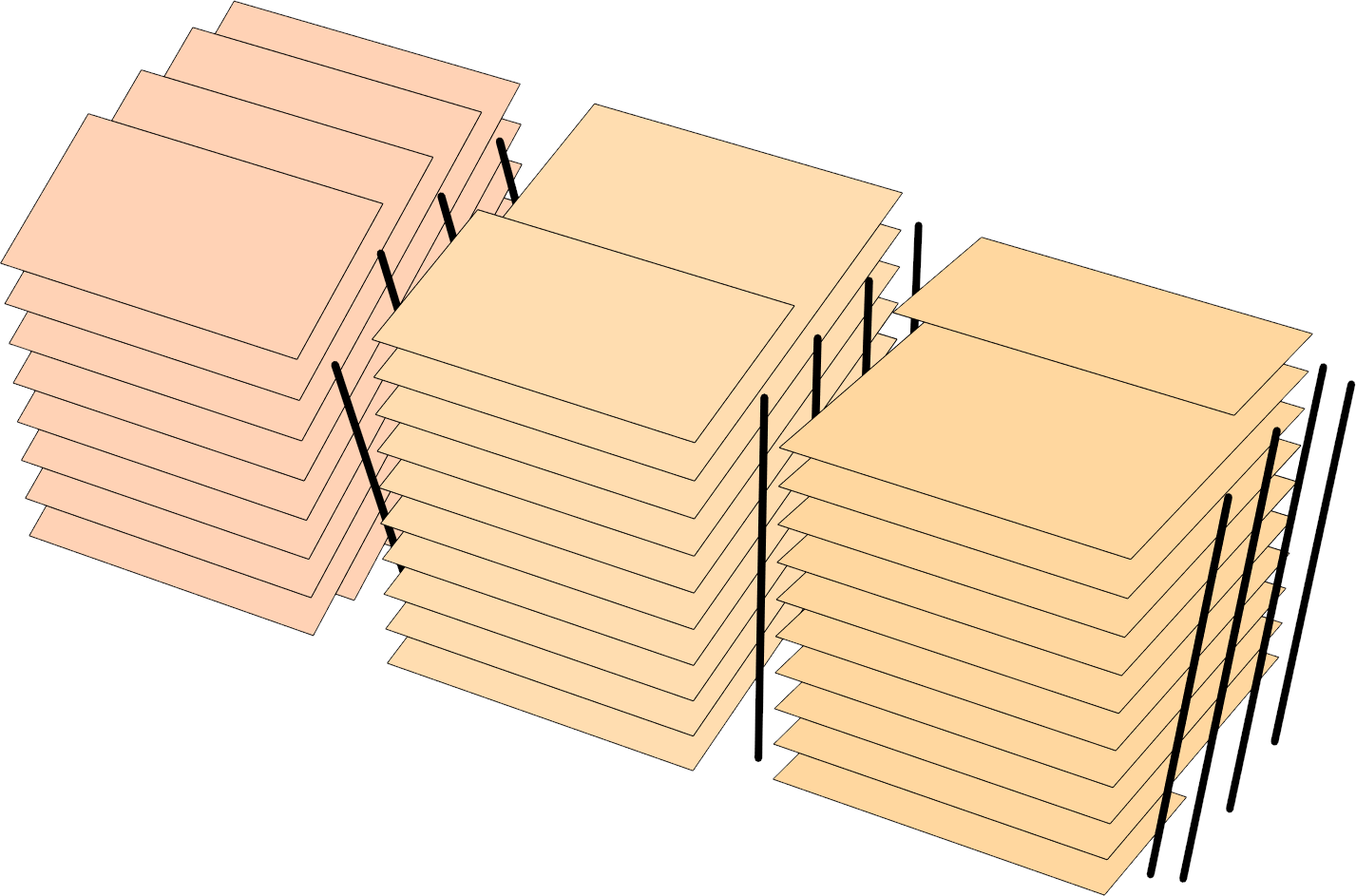}} &
\resizebox{5.2cm}{!}{\includegraphics{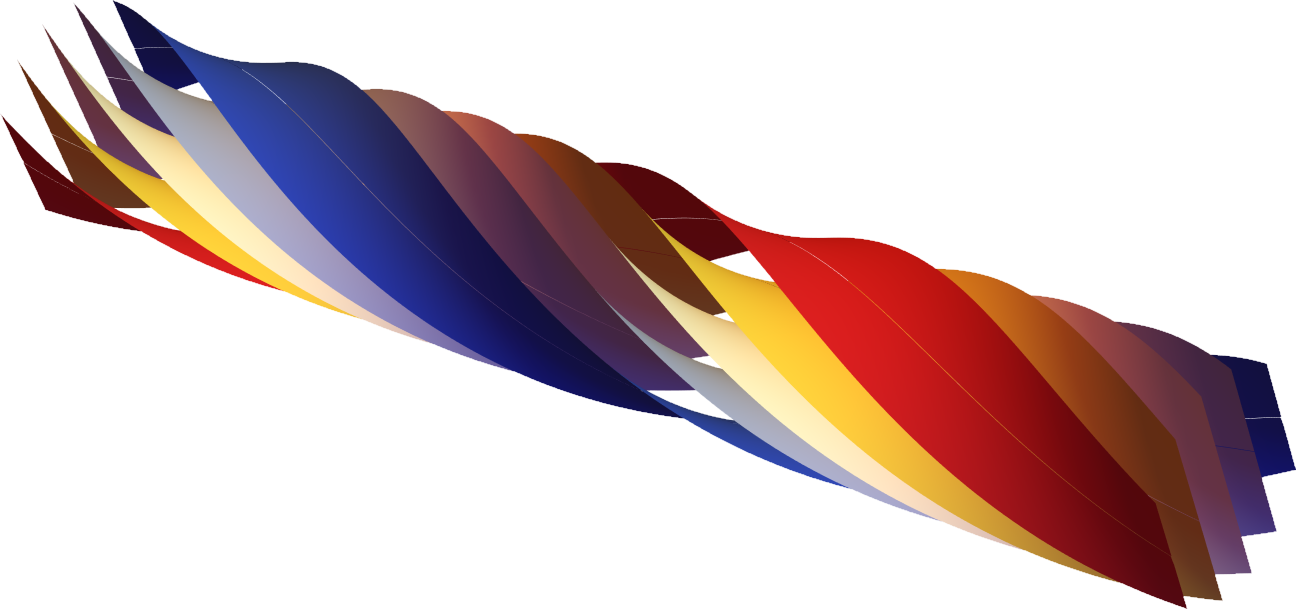}} &
\resizebox{5.2cm}{!}{\includegraphics{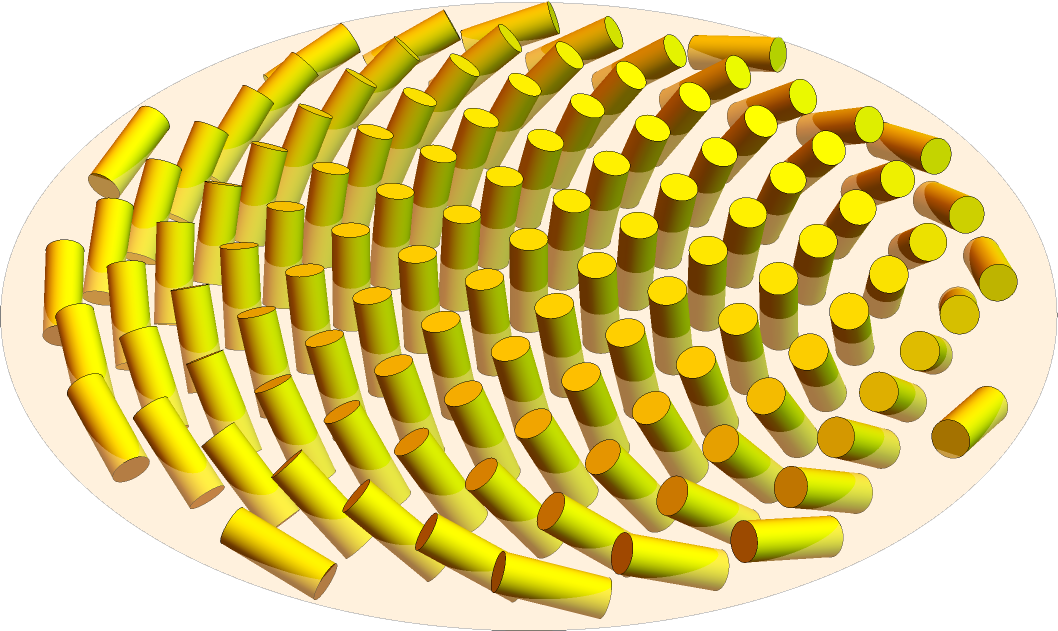}} \\
\end{tabular}
\caption{Structures involving frustration of favored twist vs.\ smectic-A order:  (a)~Twist-grain-boundary phase.  (b)~Helical nanofilament.  (c)~Colloidal membrane of chiral rods.}
\label{fig:smectic}
\end{figure}

De Gennes~\cite{DeGennes1972} pointed out an analogy between smectic-A liquid crystals and superconductors:  Twist is incompatible with smectic-A order just as a magnetic field is incompatible with superconducting order.  He suggested that a smectic-A phase might respond to favored twist in the same way that type-II superconductors respond to a magnetic field, by forming an Abrikosov lattice of defect lines.  Inspired by this analogy, Renn and Lubensky~\cite{Renn1988} proposed a model for the ``twist-grain-boundary'' (TGB) phase, shown in Fig.~\ref{fig:smectic}(a).  This structure consists of slab-like domains (``grains'') of perfect smectic-A order with no twist, separated by walls (``grain boundaries'').  Across each wall, there is a twist in the orientation of the smectic layer normal and the nematic director.  Each wall consists of an array of screw dislocations in the layer structure, and each screw dislocation is a line defect where the smectic layers are not continuous.  Hence, the nematic twist is expelled from the perfect smectic-A regions and concentrated into the screw dislocations.  The TGB phase has been identified in experiments~\cite{Goodby1989a,Goodby1989b}.  More complex variations of the TGB phase have also been studied theoretically and experimentally, including variations based on smectic-C rather than smectic-A order.

\paragraph{Helical nanofilaments}

Apart from the TGB phase, Hough et al.~\cite{Hough2009} experimentally found a different chiral phase formed by smectic layers, which they call the ``helical nanofilament'' (HNF) phase.  In this phase, the smectic layers form many twisted filaments, with the structure shown in Fig.~\ref{fig:smectic}(b).  In each filament, the layers twist around a central axis.  Each filament can be arbitrarily long along the central axis, but it is limited in size in the two transverse directions.  The structures form out of bent-core molecules, which are not chiral, but which can develop spontaneous chiral order.

Matsumoto et al.~\cite{Matsumoto2009,Matsumoto2017} developed a theory for the HNF phase, based on the favored twist arising from spontaneous chiral order.  In their theory, the filaments can be understood as long, narrow domains in which the liquid crystal satisfies two of the three requirements:  The director has twist, and the smectic layers are continuous, but the director is not exactly normal to the layers.  Rather, the director must tilt away from the layer normal, and this tilt increases away from the center of the filament.  The free energy associated with that tilt limits the size of the filament.  Hence, the theory predicts that the HNF phase forms in the ``high chirality'' limit, where the twist free energy is more important than the tilt free energy, while the TGB phase forms in the opposite ``low chirality'' limit.  The theory also shows how the layers can be connected from one filament to its neighbors in a periodic lattice; these connections might be interpreted as domain walls between the filaments.  In Section 3.2.3.1, we compare this theory with an alternative theory and with the experiment.

\paragraph{Colloidal membranes of chiral rods}

One particularly simple and well-controlled example of the frustration between chirality and smectic-A order was developed by Dogic and collaborators~\cite{Barry2009,Gibaud2012}.  They perform experiments on fd viruses, which are rod-like colloidal particles with a chiral interaction, so that they pack with a slight twist with respect to their neighbors.  Under the right solvent conditions, the viruses self-assemble into a single-layer, disk-shaped membrane with an exposed edge, illustrated in Fig.~\ref{fig:smectic}(c).  There is no twist in the interior of the disk, because it is incompatible with the favored smectic-A membrane structure.  However, there is twist at the edge of the disk, and it extends into the disk over a distance called the ``twist penetration length,'' analogous to the London penetration length for a magnetic field in a superconductor.  As the temperature is reduced, the line tension of the exposed edge decreases.  In response, the disk develops starfish-like arms, which greatly increase the length of edge, and hence more of the material becomes twisted.  These shape changes have been modeled through an elastic theory based on the competition between twist, tilt, and edge energies~\cite{Barry2009,Gibaud2012}.

\subsubsection{Splay}

Now consider a liquid crystal with parallel polar order.  As shown in Section 2.3.2, this order leads to a favored splay $\overline{S}$.  In contrast with the twist case, the favored splay is compatible with smectic-A layers.  Hence, the free energy of Equation~\ref{fsplay} becomes
\begin{equation}
F=\frac{1}{2}(K_{11}-K_{24})(\kappa_1+\kappa_2-\overline{S})^2 + \frac{1}{2}K_{24}(\kappa_1-\kappa_2)^2
=\frac{1}{2}K_{11}(2H-c_0)^2
-2K_{24}K_G
+\text{const}.
\end{equation}
This expression can be recognized as the Helfrich free energy~\cite{Helfrich1973} with a spontaneous curvature $c_0=(K_{11}-K_{24})\overline{S}/K_{11}=\lambda_\parallel\overline{\bm{P}}_\parallel\cdot\hat{\bm{n}}/K_{11}$.  The ideal local structure occurs at $\kappa_1=\kappa_2=\overline{S}/2$, so the smectic layers tend to curve into a spherical shell with radius $2/\overline{S}$.

Although there is no incompatibility between splay and smectic-A layers, we still have the nematic incompatibility between splay and 3D Euclidean geometry; it is impossible to fill space with pure constant splay.  For that reason, there are limits on the overall size of the ideal spherical shell, and on the thickness of the shell.  Hence, the achievable global phase might be a lattice of spherical shells, similar to the structure in Fig.~\ref{fig:splay}(c) but with smectic layers normal to the nematic director field.

\subsubsection{$\bm{\Delta}$ mode}

Next, suppose that a liquid crystal has octupolar order, and hence has a favored deformation $\overline{\Delta}_{ij}$.  As in the splay case, this favored deformation is compatible with the smectic-A layers.  To put it into the free energy, let us consider its tensor structure explicitly.  The favored deformation is anisotropic in the plane perpendicular to $\hat{\bm{n}}$:  outward along one axis $\hat{\bm{l}}$, and inward along an orthogonal axis $\hat{\bm{m}}$.  Hence, the favored deformation is $\overline{\Delta}_{ij}=\overline{\delta}(l_i l_j - m_i m_j)$.  Under the transformation $\hat{\bm{n}}\leftrightarrow-\hat{\bm{n}}$, we must switch the axes $\hat{\bm{l}}\leftrightarrow\hat{\bm{m}}$.

We can express the tensor structure of the actual deformation $\Delta_{ij}$ in the same basis.  We already showed that $\Delta_{ij}$ has eigenvalues of $\pm\frac{1}{2}(\kappa_1-\kappa_2)$ and $0$.  In the ideal local structure, it is reasonable that the eigenvectors of $\Delta_{ij}$ should be aligned with the eigenvectors of $\overline{\Delta}_{ij}$.  Hence, we can write $\Delta_{ij}=\frac{1}{2}(\kappa_1-\kappa_2)(l_i l_j - m_i m_j)$.

Putting these tensors into the free energy of Equation~\ref{fdelta}, we obtain
\begin{equation}
F=\frac{1}{2}(K_{11}-K_{24})(\kappa_1+\kappa_2)^2 + \frac{1}{2}K_{24}(\kappa_1-\kappa_2-2\overline{\delta})^2 .
\label{fdeltawithcurvature}
\end{equation}
This expression is different from the Helfrich free energy.  The difference is reasonable, because the Helfrich free energy assumes that a membrane is isotropic in the plane.  Here, we are considering a membrane with octupolar order, which is necessarily anisotropic.  The ideal local structure occurs at $\kappa_1=-\kappa_2=\overline{\delta}$, so the smectic layers have hyperbolic curvature.

In the literature, researchers sometimes refer to structures with favored negative Gaussian curvature or favored saddle-splay.  This is essentially the same physical concept that we are presenting here.  However, we think it is clearer to describe the concept as a favored difference of principal curvatures $\kappa_1-\kappa_2$, because the octupolar order couples directly to that difference.  In this way, we can see that a cylinder (with $\kappa_1>0$ and $\kappa_2=0$) has a lower free energy than a flat surface (with $\kappa_1=\kappa_2=0$).  By contrast, if one considers only Gaussian curvature, one might think that those two structures have equal free energy.

Like the splay deformation, the $\bm{\Delta}$ deformation is compatible with smectic layers, but it is still not compatible with 3D Euclidean geometry.  Hence, the ideal local structure cannot fill space in all directions.  In this case, the size must be limited in two directions away from a central axis, although it can grow arbitrarily long along the axis.

\paragraph{Helical nanofilaments}

Using the concept of a favored $\bm{\Delta}$ deformation due to octupolar order, we can suggest an alternative theory for the HNF phase.  Like the theory of Matsumoto et al.~\cite{Matsumoto2009,Matsumoto2017}, we suppose that the smectic layers have the shape given by $\phi_\text{loc}=x\cos q_0 z +y\sin q_0 z = \text{const}$.  Unlike that theory, we suppose that the director is aligned exactly along the local layer normal, with no tilt, so that $\hat{\bm{n}}=\bm{\nabla}\phi_\text{loc}/|\bm{\nabla}\phi_\text{loc}|$.  From that expression, we explicitly calculate all of the director deformation modes.  Along the central axis $x=y=0$ the deformation is pure $\bm{\Delta}$ mode, with $\Tr(\bm{\Delta}^2)=2q_0^2$;  near this axis it is predominantly $\bm{\Delta}$ mode.  The twist is zero, as it must be for a director field that is normal to smooth smectic layers.

Because the HNF structure has such a large component of $\bm{\Delta}$, we suggest that it may be stabilized by octupolar order of the molecules, which generates a favored $\overline{\bm{\Delta}}$.  This suggestion is essentially the same physical concept that was proposed by the experimental group of Hough et al.~\cite{Hough2009}; here we express it in terms of the four director deformation modes.  It differs from the theory of Matsumoto et al.~\cite{Matsumoto2009,Matsumoto2017}, because the mechanism does not involve chiral order and twist.  In this scenario, the helical structure of the filaments is not a sign of microscopic chiral order; it is just an accidental result of cutting smectic layers with hyperbolic curvature in a specific direction, so that they can grow arbitrarily long in the orthogonal direction, similar to work of Efrati and Irvine~\cite{Efrati2014}.

We find the mechanism based on octupolar order to be more satisfying than the mechanism based on chiral order, because octupolar order couples with $\bm{\Delta}$, which is the predominant mode near the central axis.  To be sure, either or both mechanisms may be present.  In any case, it is remarkable that the two mechanisms generate the same shape of the smectic layers.  Furthermore, the predictions of Matsumoto et al.\ for the connections between neighboring filaments may still apply with the octupolar mechanism.  These connections may favor the same helical sense for all filaments in a sample, as seen in the experiments.

\subsubsection{Bend}

Finally, consider perpendicular polar order, which induces a favored bend.  Like twist, bend is incompatible with smectic-A order, because bend induces a variation in the layer spacing.  Hence, a smectic-A liquid crystal with favored bend experiences frustration, and it might develop a complex structure with multiple domains and defects.

When de Gennes developed the analogy between smectic-A liquid crystals and superconductors~\cite{DeGennes1972}, he considered the case of bend as well as twist.  He suggested that smectic-A liquid crystal could respond to bend by forming a lattice of edge dislocations, just as it could respond to twist with a lattice of screw dislocations.  Hence, we might expect a structure analogous to the TGB phase, but with bend grain boundaries composed of edge dislocations, rather than twist grain boundaries composed of screw dislocations.  To our knowledge, this concept has not been followed up in further theoretical or experimental research; it remains a possibility for further study.

\subsection{Further constraints if layers are solid}

So far we have discussed smectic liquid crystals with fluid layers.  By comparison, now suppose that the layers are solid.  The important difference between these systems is that solid layers have a nonzero shear modulus, so that it costs elastic energy to move material points closer together or farther apart.  This elastic energy cost does not occur in systems of fluid layers, in which material can flow freely within each layer.

Sharon, Efrati, and collaborators~\cite{Klein2007,Efrati2009,Sharon2010} have recently developed a new approach for the elastic theory of thin solid films, which highlights the possibility of geometric frustration.  In this approach, the elastic energy (for large objects) or elastic free energy (for small objects with thermal fluctuations) can be written as the sum of a stretching component and a curvature component, $F=F_\text{stretch}+F_\text{curv}$.  The stretching energy $F_\text{stretch}\sim(\bm{a}-\overline{\bm{a}})^2$ depends on the difference between the actual metric tensor $\bm{a}$, which represents the distances between nearby points in the midplane of the film, and the target metric tensor $\overline{\bm{a}}$.  The curvature energy $F_\text{curv}\sim(\bm{b}-\overline{\bm{b}})^2$ depends on the difference between the actual curvature tensor $\bm{b}$ and the target curvature tensor $\overline{\bm{b}}$.

In this theoretical approach, the actual tensors $\bm{a}$ and $\bm{b}$ are related to each other through differential geometry, because they are both derived from the shape of the film.  However, the target tensors $\overline{\bm{a}}$ and $\overline{\bm{b}}$ are not necessarily related to each other; they might be determined by different physical mechanisms.  Hence, it might or might not be possible for the material to find some shape such that $\bm{a}=\overline{\bm{a}}$ and $\bm{b}=\overline{\bm{b}}$.  If not, then the targets are incompatible and the film is frustrated.  It is sometimes called a non-Euclidean film, because it cannot achieve the target metric and target curvature within 3D Euclidean space.

In the previous sections of this article, we have discussed elasticity of the nematic director field.  All of those considerations go into the curvature energy $F_\text{curv}$ for the smectic layers.  In particular, the favored director deformation modes go into the target curvature $\overline{\bm{b}}$.  If the smectic layers are fluid, then the stretching energy $F_\text{stretch}$ is zero.  Changes to the metric tensor cost no energy, because the material can flow to adapt to the new metric.  Hence, the material will take the shape favored by the curvature energy.

By contrast, if the smectic layers are solid, then the shape must be a compromise between the shapes favored by $F_\text{stretch}$ and $F_\text{curv}$.  In general, this compromise depends on the lateral size of the film, in comparison with its thickness.  If the lateral size is small, the compromise will be close to the shape favored by $F_\text{curv}$.  However, if the lateral size is large, the forces from $F_\text{stretch}$ become much larger than the forces from $F_\text{curv}$.  The material is then driven to a shape with $F_\text{stretch}\approx0$, and $F_\text{curv}$ can only select among those shapes.

\subsubsection{Lipid tubules}

Many types of amphiphilic molecules self-assemble in aqueous solution into aggregates with chiral structures, as reviewed in Reference~\cite{Spector2003}.  In some cases, these aggregates are twisted ribbons with negative Gaussian curvature, similar to the helical nanofilaments discussed above.  In other cases, the aggregates are helical ribbons with cylindrical curvature, or full cylindrical tubules that may exhibit helical markings.  As a specific example, let us consider chiral aggregates of diacetylenic phospholipids, which were found experimentally by Yager and Schoen~\cite{Yager1984,Schnur1993}.

To model the experiments, simulations of Selinger et al.~\cite{Selinger2004} found a continuous range of aggregate shapes, with narrower ribbons forming twisted structures with negative Gaussian curvature, and wider ribbons forming helical structures with cylindrical curvature.  Ghafouri and Bruinsma~\cite{Ghafouri2005} then developed an analytic theory for the shape selection, which showed a sharp transition point in terms of ribbon width and elastic constants, at which the midline begins to curve and the ribbon begins to develop cylindrical curvature.  Armon et al.~\cite{Armon2014} further extended the concept into a statistical mechanical theory for the self-assembly of these supramolecular aggregates.  In this theory, the shape is driven by a target curvature tensor $\overline{\bm{b}}$, which favors saddle-like, hyperbolic curvature.  If the ribbon is sufficiently narrow, it can achieve the target curvature by forming a twisted ribbon, with only small stretching energy.  However, as the ribbon grows wider, the necessary stretching energy becomes too large, and the ribbon buckles into a cylindrically curved structure, which has some favored curvature but not the target $\bm{b}=\overline{\bm{b}}$.  At a certain ribbon width, any further increase in width becomes energetically unfavorable, and hence the ribbon ceases to grow by widening.  It may still grow longer, and new ribbons may nucleate out of solution.

The theory of Armon et al.\ leaves an open question about the physical origin of the target curvature tensor $\overline{\bm{b}}$.  Can we understand why a supramolecular aggregate has that target curvature, in terms of the director deformation modes discussed in this article?

One answer to that question was provided many years earlier by Helfrich and Prost~\cite{Helfrich1988}.  They considered a membrane with both chiral order and anisotropy in the layer plane.  The anisotropy might occur, for example, in a smectic-C phase.  In this phase, the director $\hat{\bm{n}}$ tilts with respect to the layer normal $\hat{\bm{N}}$, and this tilt can be characterized by the projection $\bm{c}$ into the layer plane.  The director twist can then be expressed in terms of the curvature tensor $\bm{b}$ as $T=(\bm{c}\times\hat{\bm{N}})\cdot\bm{b}\cdot\bm{c}$.  With the simplifying assumption of elastic isotropy ($K_{11}=K_{22}=K_{33}=2K_{24}$), the free energy of Equation~\ref{ftwist} can be put into the form of $F_\text{curv}$, with the target curvature $\overline{\bm{b}}=\frac{1}{4}\overline{T}[(\bm{c}\times\hat{\bm{N}})\otimes\bm{c}+\bm{c}\otimes(\bm{c}\times\hat{\bm{N}})]$.  This target curvature tensor favors saddle-like, hyperbolic curvature, with principal axes at $\pm45^\circ$ from the tilt direction $\bm{c}$.  Hence, it provides one mechanism to explain the origin of the curved structures.  Several extensions of this theoretical approach have been reviewed in Reference~\cite{Selinger2001}.

We suggest that octupolar order may provide an alternative physical origin for the target curvature tensor.  For a smectic-A phase with octupolar order, the free energy is given by Equation~\ref{fdelta} or \ref{fdeltawithcurvature}.  Assuming elastic isotropy, it can be put into the form of $F_\text{curv}$, with the target curvature tensor $\overline{\bm{b}}=\overline{\bm{\Delta}}$.  This target curvature tensor favors saddle-like, hyperbolic curvature, with curvature axes aligned with the principal axes of the octupolar order.  Because this octupolar mechanism predicts the same tensor structure for $\overline{\bm{b}}$ as the chiral mechanism, it predicts the same curved structures.  As noted in Section 3.2.3.1, the octupolar mechanism implies that the observed helical shape does not occur because of microscopic chiral order, but only because of cutting the membrane in specific directions.

There are certain experimental reasons to prefer the octupolar mechanism over the chiral mechanism.  First, most lipids are chiral, but most lipids do not form tubules.  The unusual molecular feature of tubule-forming diacetylenic phospholipids is associated with kinks in the chains, which resemble bent-core liquid crystals, and not with the chiral head groups.  Furthermore, experiments of Pakhomov et al.~\cite{Pakhomov2003} synthesized an achiral version of the diacetylenic phospholipids and still found very similar helical structures, but with random handedness.  Hence, as with helical nanofilaments, lipid tubules may be another physical system where octupolar order can masquerade as chirality.

\section{LIQUID CRYSTAL ELASTOMERS}

Apart from nematic and smectic phases, liquid crystal elastomers are a third area of research on geometric frustration.  That area was reviewed by Warner~\cite{Warner2020} in this journal last year, so we will only make a few brief comments.

Liquid crystal elastomers are materials in which liquid crystal mesogenic groups are covalently bonded to a crosslinked polymer network~\cite{Warner2003,White2015}.  Because of this structure, any change in the liquid crystal order affects the shape of the polymer network, and vice versa.  The crosslinking may be weak, so that the director is a separate degree of freedom from the polymer network, or strong, so that director is locked into the network.  Some writers reserve the term ``liquid crystal elastomers'' for the case of weak crosslinking.  For the opposite case of strong crosslinking, they use the term ``liquid crystal glasses.''

In many experiments, a sample is prepared with a specified, nonuniform director configuration, which is said to be ``programmed'' or ``blueprinted.''  The sample is then heated/cooled, so that the magnitude of nematic order decreases/increases.  Because of the coupling between shape and nematic order, the material tends to contract/extend along the local director and extend/contract in the two orthogonal directions.  This change in natural shape can be expressed as a change in the 3D target metric tensor $\overline{\bm{g}}=(\lambda^2-\lambda^{-2\nu})\hat{\bm{n}}\otimes\hat{\bm{n}}+\lambda^{-2\nu}\bm{I}$, where $\lambda$ is the contraction/extension factor and $\nu$ is the Poisson ratio~\cite{Warner2020}.  For an alternative derivation of the target metric tensor, which allows for some freedom in the director, see Nguyen and Selinger~\cite{Nguyen2017}.

A sample is normally prepared as a thin film.  In that geometry, the elastic energy breaks up into a stretching energy $F_\text{stretch}\sim(\bm{a}-\overline{\bm{a}})^2$ and a curvature energy $F_\text{curv}\sim(\bm{b}-\overline{\bm{b}})^2$, as discussed in Section 3.3.  In these expressions, the 2D target metric tensor $\overline{\bm{a}}$ represents $\overline{\bm{g}}$ on the midplane, and the 2D target curvature tensor $\overline{\bm{b}}$ represents the gradient of $\overline{\bm{g}}$ across the film thickness.  If the film width is much greater than the thickness, then stretching energy is much greater than bending energy.  The Oseen-Frank free energy for director deformations is generally not important in these systems.

The stretching energy provides a very severe constraint on director deformations in a liquid crystal elastomer.  Most director configurations induce target metric tensors that are not compatible with 3D Euclidean space; the material cannot achieve the target metric tensor everywhere.  This phenomenon is a form of geometric frustration.  Because of that frustration, the sample must form a strained configuration, with internal stresses throughout the material.  One way to recognize this phenomenon experimentally is to cut the sample; a material with internal stresses will relax by changing shape after it is cut.

The frustration between director configuration and shape provides many opportunities to design liquid crystal elastomers for applications as programmable shape-changing materials.  For recent progress in this research area, we refer readers to the review by Warner~\cite{Warner2020}.

\section{CONCLUSION}

In this article, we have analyzed a wide range of modulated phases in liquid crystals, using the concept of four fundamental deformation modes.  The basic argument is that each mode couples to a type of molecular order, and conversely, each type of molecular order induces a specific deformation.  As a result, there are four types of ideal local nonuniform structure.  However, these ideal local structures are all frustrated; none can exist by itself.  In a nematic phase, certain constraints arise because of the Euclidean geometry of 3D space.  In a smectic phase, even more severe constraints arise from the layered structure.  Hence, a liquid crystal must form a more complex structure with a combination of deformation modes, or with a periodic array of defects.  Without geometric frustration there would be only a few simple modulated phases, but frustration leads to these many different possible structures.  In this sense, we might say that geometric frustration is responsible for the rich variety of modulated phases that have been found, and which continue to be found, in liquid crystals. 

\section*{DISCLOSURE STATEMENT}
The author is not aware of any affiliations, memberships, funding, or financial holdings that
might be perceived as affecting the objectivity of this review. 

\section*{ACKNOWLEDGMENTS}
This work was supported by National Science Foundation Grant DMR-1409658.

\bibliography{geometricfrustration1}
\bibliographystyle{ar-style4}

\end{document}